\documentclass[reprint,superscriptaddress,aps,pre]{revtex4-1}

\usepackage{graphicx}
\usepackage{amssymb}
\usepackage{amsmath}
\usepackage{epsfig}
\usepackage{chemarr}
\usepackage{url}
\usepackage{natbib}
\usepackage{dcolumn}
\usepackage{bm}
\usepackage{color}
\usepackage{braket}

\def\be{\begin{equation}}
\def\ee{\end{equation}}
\def\bmu{\begin{multline}}
\def\bea{\begin{eqnarray}}
\def\eea{\end{eqnarray}}

\def\p{\partial} 

\def\f{\frac}

\def\l{\left(}
\def\r{\right)}

\def\grad{\vec{\nabla}}

\begin{document}

\title{Physical constraints on epistasis}

\author{Kabir Husain}
\affiliation{James Franck Institute and the Department of Physics, University of Chicago, Chicago IL, USA}
\author{Arvind Murugan}
\email{amurugan@uchicago.edu}
\affiliation{James Franck Institute and the Department of Physics, University of Chicago, Chicago IL, USA}
\begin{abstract}
Living systems evolve one mutation at a time, but a single mutation can alter the effect of subsequent mutations. The underlying mechanistic determinants of such epistasis are unclear. Here, we demonstrate that the physical dynamics of a biological system can generically  constrain epistasis. We analyze models and experimental data on proteins and regulatory networks. In each, we find that if the long-time physical dynamics is dominated by a slow, collective mode, then the dimensionality of mutational effects is reduced. Consequently, epistatic coefficients for different combinations of mutations are no longer independent, even if individually strong. Such epistasis can be summarized as resulting from a global non-linearity applied to an underlying linear trait, i.e., as global epistasis. This constraint, in turn, reduces the ruggedness of the sequence-to-function map. By providing a generic mechanistic origin for experimentally observed global epistasis, our work suggests that slow collective physical modes can make biological systems evolvable.
\end{abstract}

\maketitle

Function in biology arises from the concerted, collective action of individial components. A mutation in any one component can alter the collective state of the system, modulating the effect of later mutations. Multiple mutations can therefore have effects distinct from the sum of their parts. Such epistasis, or non-additivity (Fig.~\ref{fig:Outlook}a), is found across scales of biological organisation: from individual proteins \cite{Starr2016,Adams2019,Domingo2019}, to intracellular genetic \cite{Macia2012,RojasEchenique2019} and metabolic \cite{Segre2005,Bajic2018} networks, to complex ecosystems of many species \cite{Gould2018,Grilli2017}. Apart from being difficult to measure experimentally, epistasis can also hinder the natural evolution of a biological system, by making sequence-to-function maps highly rugged \cite{Kauffman1989}, with many local fitness peaks that can trap evolutionary trajectories and thus make evolution more contingent, less predictable, and less reversible\cite{Starr2016,Domingo2019,Starr2017,Starr2018}.

However, recent analyses of deep mutagenesis data \cite{Otwinowski2018a,Otwinowski2018,Sailer2018} have suggested that epistasis in some real biological systems may instead be highly constrained. Known as global or non-specific epistasis\cite{Starr2016}, such epistasis takes the form of a global non-linearity distorting an underlying additive trait. 
For example, mutations themselves might be effectively additive in determining the folding energy $E$ of a protein, but folding stability, quantified by a non-linear function $f(E)$ of energy (e.g.,  $f \sim \exp(-E/k_B T)$), will nevertheless show strong epistasis. \cite{Otwinowski2018}. 
In this case, only a few independent numbers -- and thus only a few measurements -- suffice to specify all other mutational interactions. In addition, the evolutionary consequences of global epistasis are also different: the fitness landscape is expected to be less rugged and easier to navigate \cite{Kryazhimskiy2014}.

What are the mechanistic determinants of constrained epistasis \cite{Domingo2019, Lehner2011,DePristo2005,Sailer2017b,Otwinowski2018}? That is, what kinds of underlying physical interactions or mechanisms give rise to fully rugged, or `specific' \cite{Starr2016}, epistasis -- with accompanying unpredictability and complexity -- versus interactions that give rise to a simplified global epistasis instead? One obvious answer is that strong and direct physical interactions give rise to specific epistasis, while non-interacting or weakly interacting sites (e.g., additive contributions to free energy  \cite{Otwinowski2018}), with a non-linearity introduced elsewhere (e.g., in measurement, or in relating function to fitness \cite{Lehner2011}), leads to global epistasis. While weak interactions surely lead to global or no epistasis, many cases of interest in biology involve strong physical interactions, as in interactions between residues in the compact core of a protein structure. 

Here, we argue that global epistasis can generically arise in biological systems, if their long-time \textit{physical} dynamics -- their response to transient perturbations -- is described by only a few collective degrees of freedom. These include proteins with a few `soft' mechanical or vibrational modes, as well as gene regulatory networks whose expression levels relax primarily along a few slow collective modes. We show that these systems exhibit a stereotyped response to mutations, with combinations of mutations forced to interact only through the few slow collective physical modes.  This, in turn, implies global epistasis, which, in a small perturbation approximation, we diagnose by a reduced rank of the epistatic matrix: that is, the matrix of $\mathcal{O}(N^2)$ pairwise epistatic interactions between $N$ mutations is specified entirely by $\mathcal{O}(N)$ numbers. The corresponding evolutionary fitness landscapes is found to be less rugged and easier to navigate, suggesting that `evolvability' can be selected for by selecting for a separation of biophysical timescales. 

Notably, while our arguments rely on testable assumptions about mutations -- which we confront with structural data on proteins and proteomic data in \textit{E. coli} -- our analysis applies broadly to \emph{any} fitness function in such systems, and might also apply to other non-mutational, high-order interacting systems in biology, such as drug interactions \cite{Yeh2006,Chait2007,Wood2012} or ecological networks \cite{Grilli2017,Sanchez-Gorostiaga2018}. Overall, our results suggest a physical origin for constrained evolutionary epistasis, and rationalizes the occurrence of slow, collective modes in proteins and biological networks.

\section*{Results}

\begin{figure}
\includegraphics[width=0.85\linewidth]{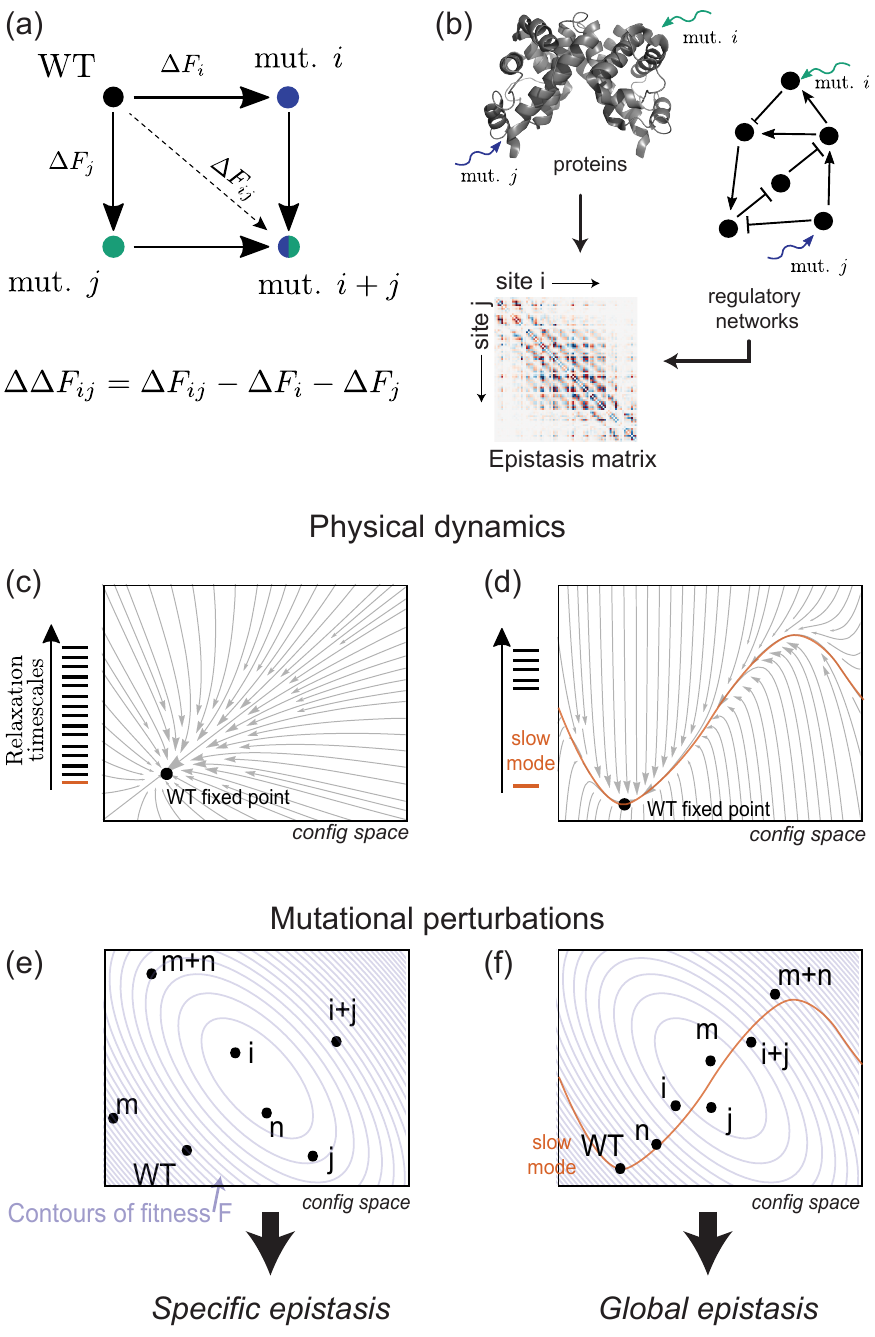}
\caption{ \label{fig:Outlook} \textbf{Slow physical modes can constrain mutational interactions.} 
\textbf{a} The non-additivity of two mutations $i$ and $j$ on fitness $F$ defines the epistatic coupling $\Delta \Delta F_{ij}$; \textbf{b} epistatic couplings are found at many scales, from within a single protein to across the genome. 
\textbf{c, d} We contrast epistasis in systems with qualitatively different \emph{physical} dynamics, e.g., the relaxational dynamics of protein structure or expression levels in a regulatory network. Some biological systems \textbf{d} naturally show a large gap in the spectrum of timescales of such relaxation modes; i.e., some relaxation modes (orange) are much slower than others. 
\textbf{e} We argue that without slow modes as in \textbf{c}, the effect of mutations is idiosyncratic in the high-dimensional configuration space of the system (e.g. the molecular coordinates of all atoms in a protein). 
\textbf{f} In contrast, with a slow mode as in \textbf{d}, we argue that the effect of mutations is mostly confined to the low dimensional subspace (orange) defined by the slow mode. Consequently, while epistatic coupling for each pair $(i,j)$ or $(m,n)$ can still be as strong as in \textbf{e}, the couplings are now related to each other through the non-linearity of the fitness function $F$ (purple) along the slow mode (orange), leading to global epistasis. 
}
\end{figure}

We present our results in terms of distinct models of protein and regulatory network function. In each, we find that if there is a slow dynamical mode (Fig.~\ref{fig:Outlook}c,d), then mutations only perturb the biological system along a low-dimensional contour in an otherwise high-dimensional configuration space (Fig.~\ref{fig:Outlook}e,f). As a result, epistasis is global, arising from the non-linearity of fitness along that slow mode.

\subsection{Soft Modes in a Protein Constrain Epistasis}

We begin with proteins. The three-dimensional arrangement of atoms in folded proteins forms the physical basis of protein function (i.e. binding to a ligand, or the catalysis of a reaction). A mutation in sequence -- i.e. the replacement of one amino acid for another -- leads to a deformation in the folded structure (or in an ensemble of folded structures), and a concomitant change in function (e.g. an altered $K_d$).

\begin{figure*}
\includegraphics[width=\textwidth]{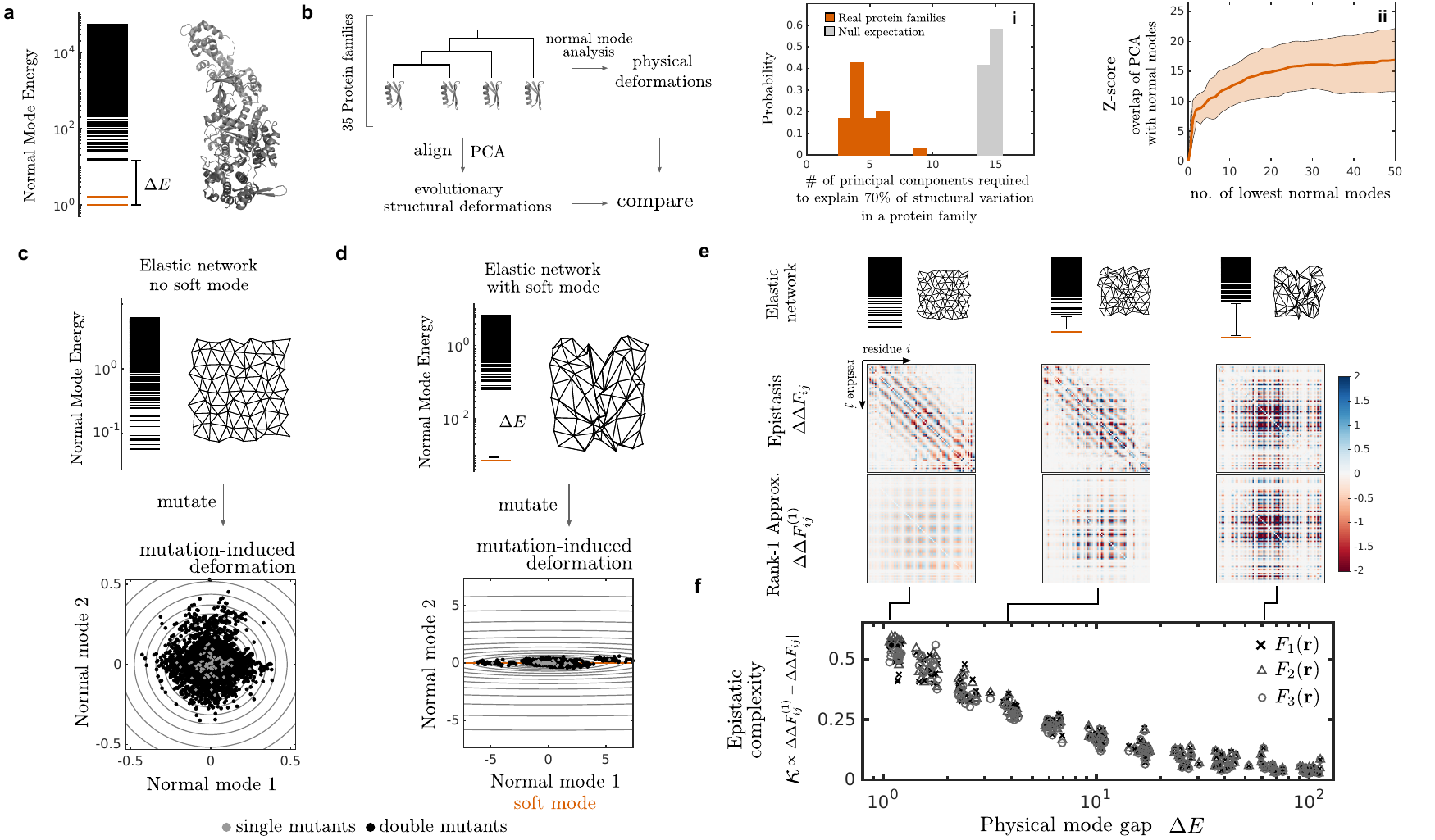}
\caption{  \textbf{Soft modes in proteins constrain mutation-induced structural deformations and thus simplify epistasis.} 
\textbf{a} Many proteins exhibit one, or a few, soft modes of deformation (orange); quantified by a gap $\Delta E$ in their normal mode spectrum (shown for Myosin V, PDB ID: 1W7J).
\textbf{b} Variations in protein structure across a protein family (due to mutations) are (i) low dimensional (4-5 PCA components); further, (ii) these components have highest overlap with the softest normal modes of a given protein in that family. (PCA components and Z-scores for overlap reproduced from \cite{Leo-Macias2005})
\textbf{c, d} Simulated mutation-induced deformations in an elastic network model without (\textbf{c}) and with (\textbf{d}) a soft mode, shown in the space spanned by the two lowest normal modes; mutations in the background of a soft physical mode deform the protein mostly along that soft mode (energy contours of WT shown in gray).
\textbf{e} We perform double mutant cycles in a family of ENMs with varying mode gap $\Delta E$ (i.e. the ratio of the two lowest normal mode energies), measuring three arbitrarily chosen fitness functions $F(r)$ of structure (symbols, see SI). The resultant epistatic matrices $\Delta \Delta F_{ij}$, and their rank-1 approximations, $\Delta \Delta F_{ij}^{(1)}$, are shown.
\textbf{f} The epistatic complexity, defined as the median residual $\Delta \Delta F_{ij}^{(1)} - \Delta \Delta F_{ij}$, decreases as the physical mode gap $\Delta E$ increases.
}
\label{fig:Proteins}
\end{figure*}

Interestingly, many proteins are known to be constrained in their \textit{physical} deformations -- i.e., in their response to applied forces due to e.g. ligand binding. Fig. \ref{fig:Proteins}a shows an example from a normal mode analysis, which decomposes the physical motions of a protein into a spectrum of orthogonal modes (also known as vibrational modes) \cite{Bahar2010}. A few `soft' collective motions are found to be of lower energy and thus easier to actuate than other stiffer motions. These collective motions have been extensively studied by molecular dynamics techniques \cite{Brooks1983}, compared against NMR and time-resolved X-ray crystallography \cite{Bahar2010,Hekstra2016}, and have been proposed as the mechanical basis for protein function, e.g. ligand specificity and allosteric communication \cite{Rocks2017,Dutta2018,Wodak2019,Yan2017,Rivoire2019}.

Notably, several studies have found that \textit{physical} soft modes constrain \textit{mutation-induced} deformations of a protein to be low dimensional. For e.g., in Fig.\ref{fig:Proteins}b, we summarize results\cite{Leo-Macias2005} from a study that compared the variations of protein crystal structure across a protein family, due to mutations, to the geometry of normal modes of single proteins in that family. For 35 different protein families, \cite{Leo-Macias2005} found that the structural variations across a given family are relatively low-dimensional (4 or 5 dims) and are mostly contained in the subspace spanned by the lowest energy physical modes of any protein in that family. Other studies have performed similar but detailed analyses for particular protein families such as globins\cite{Echave2010}, the Ras superfamily\cite{Raimondi2010}, DHFR \cite{Rod2003}, HIV-1 protease \cite{Yang2008}, and T4 lysozyme \cite{Sinha2002}. 

We argue that this empirical relationship between \textit{physical} soft modes and \textit{mutation-induced} deformations is to be expected mathematically. Assuming only local physical interactions between residues, we derive the structural deformation $\delta_i \mathbf{r}$, induced by a mutation at residue $i$, in terms of the physical modes of the protein. In the limit of small perturbations, we find (see SI)
\be \label{eq:mutationInducedDeformation}
\delta_i \mathbf{r} \approx \mathcal{H}_0^{-1} \grad \delta_i E (\mathbf{r}_0) \approx \f{1}{\lambda_0}  \l \hat{\mathbf{v}}_0\cdot\grad\delta_i E \r \, \hat{\mathbf{v}}_0 +  \mathcal{O} \l\frac{1}{\lambda_1} \r  
\ee
\noindent where $\mathcal{H}_0$ is the Hessian of the wildtype protein's folding energy about its folded structure $\mathbf{r}_0$, $\delta_i E$ is the perturbation in folding energy induced by the mutation. $\lambda_0$ and $\hat{\mathbf{v}}_0$ are the energy and geometry of the softest mode; while $\lambda_1$ is the energy of the next softest mode.

\indent That is, in the presence of a soft physical mode, each mutation deforms the protein along that one mode $\hat{\mathbf{v}}_0$, and mutated structures populate a one-dimensional contour in the high-dimensional space of possible structures. 

\indent This stereotyping of mutation-induced deformations, Eq. \ref{eq:mutationInducedDeformation}, is sufficient to simplify epistasis. Consider a fitness function, $F$, representing, say, the binding affinity for a ligand, or catalytic efficiency -- we place no constraints on the fitness function, supposing only that it can be written as a function of protein structure: $F(\mathbf{r})$.  As shown in the SI, in general the epistatic co-efficient between mutations $i$ and $j$ is $\Delta \Delta F_{ij} =  (\delta_i \mathbf{r} \cdot \nabla)  (\delta_j \mathbf{r} \cdot \nabla)  F(r_0)$, which reduces in the presence of a single soft mode to

\be \label{eq:lowRankEpistasis}
\Delta\Delta F_{ij} \propto \f{1}{\lambda_0^2} \, \theta_i \theta_j + c_{ij},
\ee i.e., a rank-1 matrix $\theta_i \theta_j$ (here, $\theta_i = \hat{\mathbf{v}}_0\cdot\grad\delta_i E$), and a sparse matrix $c_{ij}$ that is non-zero only for residues $i,j$ in physical contact. Notably, rank-1 matrices are highly constrained, with much fewer independent numbers than a full-rank matrix, and can therefore be reconstructed by fewer measurements. If multiple deformation modes are soft, say $k$ soft modes, the predicted rank of $\Delta\Delta F_{ij}$ is $k$. Similar observations about epistasis have been made for functions related to protein allostery \cite{Dutta2018,Eckmann2019}.

Intuitively, this reduction in complexity is due to a dimensional reduction in mutational effects. In principle, the fitness $F(\mathbf{r})$ is a function of all molecular co-ordinates, and thus can be extremely complex. However, if mutations can sample only a limited set of deformations, as in Eq. \ref{eq:mutationInducedDeformation}, the fitness landscape depends only on those few collective coordinates; see Fig.~\ref{fig:Outlook}d. Mathematically, in the background of a soft mode, the fitness function is simply $F(\{s_i\}) = g( \sum_i \theta_i s_i)$ where $g(\phi)$ is the non-linearity of the fitness function $F(\mathbf{r})$ along the soft mode, $s_i = 0,1$ is the genotype, and $\sum_i \theta_i s_i$ is the total mutation-induced deformation along the soft mode. This form, with an overall non-linearity applied to an underlying linear trait, is that of global epistasis \cite{Otwinowski2018a}. Moreover, expanding any global epistatic function $F(s_i) = g( \sum \theta_i s_i)$ generates strong yet low-rank epistasis of all orders: the 2nd order term is rank-1 (as in Eq. \ref{eq:lowRankEpistasis}) and higher order terms are similarly constrained.

\indent The above result applies for an infinitely large gap $\Delta E$ in the normal mode spectrum. We sought to test our results for finite gaps by simulating elastic network models (ENM) of proteins with varying gap $\Delta E$. ENMs have been widely used to analyze ligand binding\cite{Tobi2005,Bakan2009,Wang2019} and allostery \cite{Rocks2017,Yan2017,Dutta2018}, and are believed to robustly capture the low-frequency motions of many proteins  \cite{Tirion1996,Bahar2010}. Introducing mutations in the family of ENMs (see SI for details), we find that mutations in the background of a soft mode result in deformations along that one mode (Fig.~\ref{fig:Proteins}c,d), confirming Eq. \ref{eq:mutationInducedDeformation}. We then performed double-mutant cycles in our simulated networks, measuring epistasis for three arbitrarily chosen fitness functions $F(\mathbf{r})$ detailed in the SI. Fig.~\ref{fig:Proteins}e presents the epistatic matrix $\Delta\Delta F_{ij}$, compared against a rank-1 approximation.

We find that as the gap $\Delta E$ in the \textit{physical} mode spectrum increases, the \textit{epistatic} `complexity' (here, deviation from rank-1) decreases (Fig.~\ref{fig:Proteins}f), confirming Eq. \ref{eq:lowRankEpistasis}. 

\subsection{Slow Modes in Molecular ensembles}
An alternative picture of proteins emphasizes the statistical ensemble over a set of discrete configurations $a= 1, \ldots S$; e.g., these configurations could correspond to small structural re-arrangements of the protein, such as broken hydrogen bonds, salt bridges or shifted helices \cite{Best2006}. The WT sequence leads to occupancies $\psi_a^{\text{(WT)}}$ of these different states $a= 1, \ldots S$, while mutants can alter the energies of these states and thus result in a different set of occupancies $\psi_a^{\text{(mut.)}}$. Fitness $F(\{\psi_a\})$ is determined by these occupancies; such molecular ensembles can generically lead to epistasis \cite{Sailer2017b}. 

If we perturb the steady state occupancies $\psi_a^{\text{(WT)}}$ to $\psi_a^{\text{(pertub.)}}$ by some transient physical perturbation (e.g., raising and lowering temperature), $\psi_a^{\text{(pertub.)}}$ will relax back to $\psi_a^{\text{(WT)}}$ in a way determined by the network of kinetic barriers between states. Such a relaxation process is described by a master equation,
\begin{equation}
    \frac{d\psi_a}{dt} = \sum_b M_{ba} \psi_b - \sum_b M_{ab} \psi_a
\end{equation}
where $M_{ab},M_{ba}$ are kinetic rates of transition between states $a,b$. The lowest eigenvalue of $M$ is always zero, reflecting conservation of probability and the corresponding eigenvector gives the steady-state distribution. The next eigenvalue $\lambda_1$ and corresponding eigenvector of $M$ determine the slowest relaxation mode in the system. If this mode is much slower than the subsequent modes, i.e., if $|\lambda_1| \ll |\lambda_p|,  p>1$, then the system is characterized by a slow relaxation mode $\lambda_1, \mathbf{\hat{v}_1}$. For example, such a mode can physically correspond to a meta-stable state (or more generally, a meta-stable ensemble) that generic ensembles quickly relax to; the meta-stable ensemble then slowly relaxes to the steady state distribution. 

In the presence of such a slow mode, our earlier analysis follows through. Mutations shift the ensemble in a low-dimensional stereotyped manner. Consequently, epistasis is of the global type and in a perturbative expansion, we find that the 2nd order epistatic term $\Delta \Delta F_{ij}$ is rank-1; see SI for detailed calculations.

\subsection{Slow Modes in Regulatory Networks}

\begin{figure*}
\includegraphics[width=\textwidth]{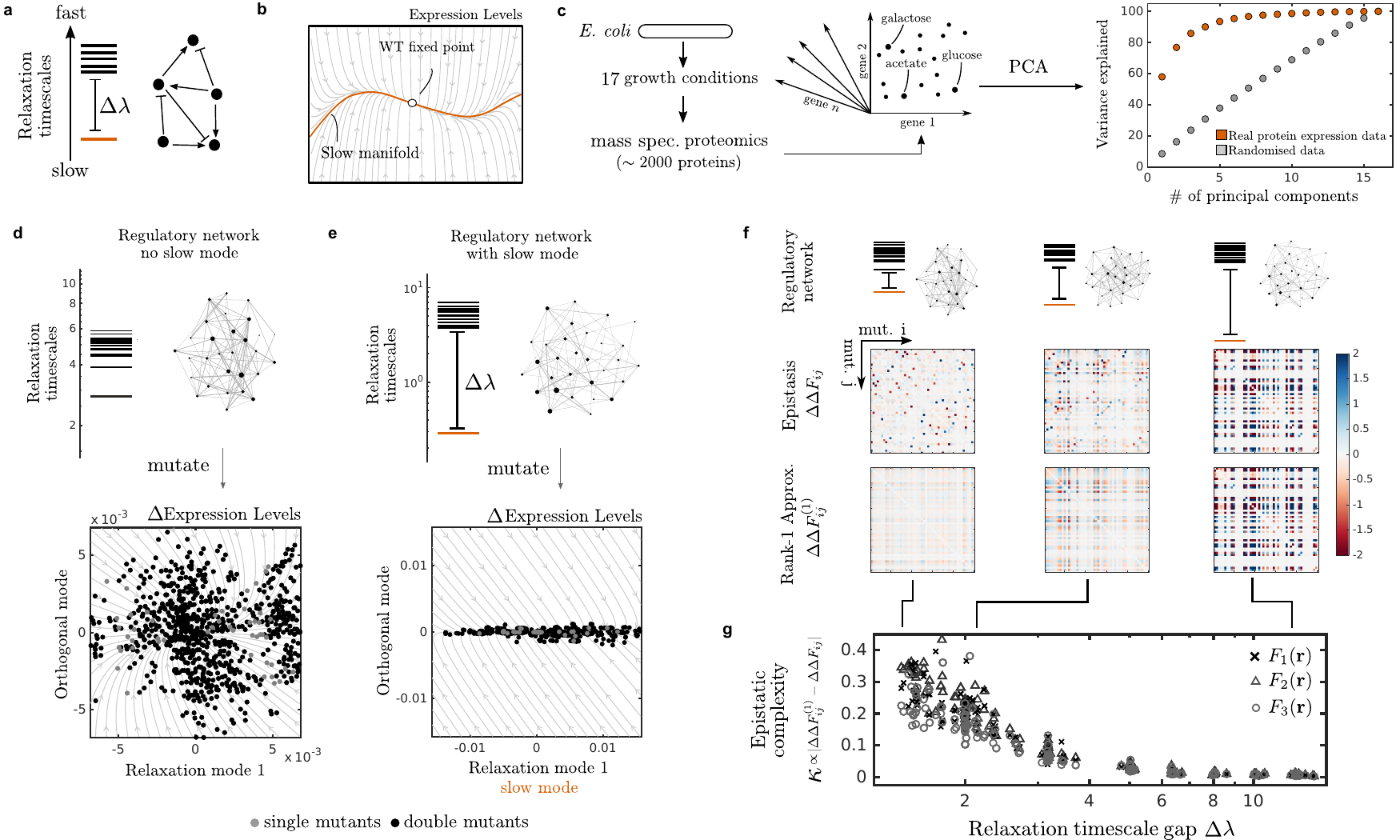}
\caption{ \label{fig:Networks} 
\textbf{Slow modes in regulatory networks constrain perturbation of expression levels and simplify epistasis.}
\textbf{a} Regulatory genetic or metabolic networks exhibit a spectrum of relaxation timescales when perturbed.  If a few modes (orange) are significantly slower than others, quantified by a gap $\Delta \lambda$ in the rate spectrum, most perturbations quickly relax to that low dimensional manifold (orange in \textbf{b}) defined by those modes; consequently, expression levels, if measured on longer timescales, are mostly contained within these few slow modes.
\textbf{c} Data from \cite{Schmidt2016,Barenholz2016} suggests that the changes in the \textit{E. coli} proteome (assessed by PCA,orange), measured across 17 different growth conditions, are contained within a low-dimensional manifold (i.e. only 3-4 principal components; gray datapoints obtained by scrambling measured expression levels across growth conditions).
\textbf{d, e} We simulate the effect of mutations in a family of regulatory networks (only top 50\% of links shown for clarity, node size represents WT expression level) without (d) and with (e) a slow mode, showing changes in expression in the subspace spanned by the lowest mode and an orthogonal direction. In the presence of a slow mode, mutations shift expression levels mostly along the slow mode.
\textbf{f} Epistasis $\Delta \Delta F_{ij}$, and its low-rank approximation $\Delta \Delta F_{ij}^{(1)}$, from simulated double mutant cycles in regulatory networks, for three randomly generated fitness functions (see SI).
\textbf{g} The epistatic complexity $\kappa \propto \vert \Delta \Delta F_{ij}^{(1)} - \Delta\Delta F_{ij} \vert$, reduces as the gap in the relaxation spectrum, $\Delta \lambda$, increases.
}
\end{figure*}

Informed by our protein results, we wondered if similar constraints on epistasis may be at play at higher levels of biological organization. In fact, epistasis has been widely investigated at the genomic, metabolic and even  ecological levels, often through the lens of  the topology of the underlying interaction networks \cite{Macia2012,Domingo2019, Segre2005, Bajic2018, Gould2018,RojasEchenique2019}.

Instead of network topology, we focus here on the underlying dynamics. The dynamics of these biological networks can be characterized by relaxation modes and an accompanying spectrum of timescales (Fig. \ref{fig:Networks}a). For e.g., when a cell is transiently perturbed by changing external conditions (e.g., available nutrients or stressors), the initial perturbation to gene expression levels excites a combination of modes. However, modes with fast relaxation timescales decay rapidly and the response on longer timescales is dominated by the slowest modes.

If a few modes are significantly slower than all other modes, most perturbations will result in shifts in expression levels along the same low dimensional subspace, as measured on longer timescales. Such low dimensionality of expression levels has been reported in proteomics experiments \cite{Schmidt2016,Barenholz2016} (reproduced in Fig.\ref{fig:Networks}c), analyzed in \cite{Kaneko2015,Furusawa2018,Kaneko2018,Sato2019}, and have also been discussed in the context of metabolic regulation of the proteome \cite{You2013,Erickson2017}.

Just as a soft mode in a protein stereotypes its response to mutations, a slow manifold in a regulatory network stereotypes its response to mutations or sustained external perturbations. Mathematically, if $\delta_i \mathbf{f}$ represents, say, a change in some regulatory rates due to a mutation $i$, the shift in steady state expression levels $\delta \mathbf{n}$ may be written as:
\be
\delta_i \mathbf{n} \approx \mathcal{J}_0^{-1} \, \delta_i \mathbf{f} \propto \f{1}{\lambda_0} \, \theta_i \, \hat{\mathbf{v}}_0 + O\l \f{1}{\lambda_1} \r
\ee
\noindent where $\mathcal{J}_0$ is the Jacobian of the dynamical equations of unperturbed network near its steady state, $\f{1}{\lambda_0}$, $\hat{v}_0$ are the timescale and direction of the slow manifold (computed as eigenvalues and eigenvectors of $\mathcal{J}_0$), and $\theta_i = \delta_i \mathbf{f} \cdot \hat{\mathbf{v}}_0$. Thus, just as with proteins, mutations in the background of a slow manifold perturb the network along a characteristic direction. 

Correspondingly, as in Eq. \ref{eq:lowRankEpistasis} for proteins, we expect that the epistasis is explained by a global non-linearity and, in a small perturbation expansion, the epistatic matrix is low-rank; see SI for derivation. 

\indent To test this result, we construct a generic, statistical family of models that are representative of biochemical networks. To be concrete, we consider a family of genetic networks, in which each node represents a gene and edges correspond to, e.g., transcriptional regulation. If $n_a$ is the expression level of gene-product $a$, the dynamics of the network may be written as:
\be \label{eq:enzymeNetwork}
\f{\p}{\p t}n_a = \sum_b k_{ab} \f{n_b}{n_b + K_{ab}} - \lambda_a \, n_a
\ee
\noindent where we have chosen Michaelis-Menten kinetic forms for the regulatory terms, with $k_{ab}$ and $K_{ab}$ parameterizing the regulation $b\to a$, and $\lambda_a$ are decay-rates for each gene product due to dilution or degradation. For simplicity, we consider only positive regulation, and suppose that each mutation affects the kinetic parameters $k_{ab}$ and $K_{ab}$ for a particular link $b \to a$ (see Fig. \ref{fig:Networks}d,e).

\indent We perform simulated double mutant cycles, varying the gap $\Delta \lambda$ in the spectrum of relaxational timescales, and measure fitness $F(\mathbf{n})$ with three randomly generated fitness functions (see SI). We again find that epistatic complexity $\kappa$, defined as deviation from rank-1, reduces with an increasing gap $\Delta \lambda$, Figs \ref{fig:Networks}f,g. 

Thus, we find that even without tuning the motifs and topology of the network \cite{Segre2005,RojasEchenique2019}, slow collective modes in strongly interacting biological networks lead to constrained, low-rank epistasis.  

\subsection{Evolutionary consequences}
Our work here proposes that slow collective physicals modes constrain the epistatic coefficients $\Delta\Delta F_{ij}$ measured in deep mutational screens. Such constrained, or global, epistasis might also affect the ruggedness of sequence-to-function maps and thus have evolutionary consequences \cite{Kryazhimskiy2014}. To test this, we considered a family of epistatic matrices $\Delta\Delta F_{ij} = (1-\kappa) \, \theta_i\theta_j + \kappa \, J_{ij}$, parameterising fitness landscapes that interpolate between those with complex epistasis as seen with no physical mode gap (when $\kappa = 1$) and those with simplified epistasis as elicited with a large mode gap (when $\kappa = 0$). Here $J_{ij}$ is a randomly generated full-rank matrix, while $\theta_i\theta_j$ is a random rank-1 matrix.

\begin{figure}
\includegraphics[width=0.95\linewidth]{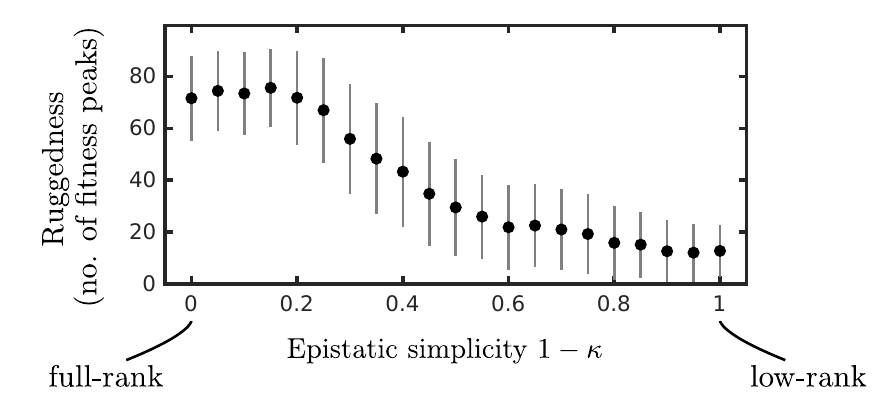}
\caption{ \label{fig:LessRugged} \textbf{Low-rank epistasis makes the sequence-to-function map less rugged.} We simulated adaptive walks on fitness landscapes of varying epistatic complexity, interpolating between low-rank and full-rank epistasis. The number of distinct fitness maxima found, a measure of the ruggedness of the fitness landscape, decreases as epistasis is tuned to be low-rank (error bars show standard deviation over 100 generated fitness landscapes; see SI for details). In conjunction with Figs. \ref{fig:Proteins}f, \ref{fig:Networks}f, these results show that slow dynamical modes in proteins and regulatory networks lead to low-rank epistasis and consequently less rugged sequence-to-function maps.
}
\end{figure}

\indent We simulated adaptive walks on fitness landscapes corresponding to different $\kappa$. As shown in Fig. \ref{fig:LessRugged}, for small $\kappa$, adaptive walks discover and get trapped at many more local maxima than for large $\kappa$, even though the size of individual epistatic couplings $ \Delta\Delta F_{ij}$ are equally large for all $\kappa$. See SI for more details.

When combined with Figs.~\ref{fig:Proteins}f,~\ref{fig:Networks}g, our results suggest a striking relationship between \textit{physical} mode gap $\Delta E$, the complexity $\kappa$ (or equivalently, rank) of the resulting \textit{mutational} epistatic matrix and consequently the ruggedness of \textit{evolutionary} fitness landscapes.

\section*{Discussion}

\indent We have argued that the physical dynamics of a biological system can easily and generically constrain its epistatic complexity. We considered proteins and regulatory networks whose response to perturbations is dominated by a few slow collective modes on longer timescales. We argued, with support from analyses of existing data sets, that such slow collective modes can effectively limit accessible deformations to a low dimensional subspace of a high-dimensional configuration space (e.g., atomic structure or ensemble occupancies for proteins, expression levels for gene regulatory networks). Correspondingly, epistasis is predicted to be global (i.e., a global non-linearity applied to an underlying linear trait) and the pairwise epistatic matrix is predicted to be low rank.

Our theory relies on experimentally testable assumptions about mutation-induced deformations and their relationship with physical modes. For instance, a conservative estimate of the mode gap required to give global epistasis gives a scaling with system size $N$ as $\sqrt{N}$. For a protein with $\sim 100$ amino acids or a regulatory network with a $100$ genes, this requires a mode-gap of $\lambda_{1}/\lambda_{0} \approx 10$. This is in line with estimates for mode gaps in real proteins \cite{Togashi2007}. We also presented evidence from the literature for how such slow modes constrain mutational-deformations in proteins, which typically compare experimentally determined mutation-induced structural deformations to computed normal modes \cite{Sinha2002, Echave2010}, and generally find a non-trivial overlap between the two \cite{Leo-Macias2005}.

\indent A deeper test of our ideas requires a detailed experimental comparison between physical and mutational experiments. For instance, in proteins a direct experimental characterization of soft modes is to be compared against measured epistasis via deep mutational scans. Data of both types are currently coming online. Deep mutational scans have been performed for a few proteins \cite{Olson2014,Salinas2018,Diss2018} (and in fact, shown to be consistent with global epistasis \cite{Otwinowski2018a}). Meanwhile, electric-field stimulated X-ray (EFX) experiments have directly probed the soft modes of a protein, by applying an electric field and capturing the deformations of a protein by time-resolved X-ray crystallography on the $100$ ns timescale \cite{Hekstra2016}. Similarly, in the context of gene regulation, one should correlate the effects of mutagenesis with physiological perturbations such as nutrient shifts \cite{Schmidt2016, Erickson2017}, as well as with measurements of genetic epistasis.

Notably, while our results rely on the nature of mutation-induced deformations, we make few assumptions about the fitness function itself; any fitness function of coordinates in which a slow mode dominates is predicted to show global epistasis. Further, we showed that the relevant slow modes can exist in diverse spaces - structure (quantified by normal mode energies), statistical ensembles (eigenvalues of a master equation) or non-linear regulatory networks (Lyapunov exponents). Note that the latter two systems may not have any equilibrium energy function, with Lyapunov exponents playing the role of normal mode energies.

\indent More broadly, our work highlights the need to understand the connection between the underlying dynamics of a system and its observed epistasis \cite{Sailer2017b,Chure2019}. 
Traditionally, strong specific epistasis is often interpreted as evidence of direct interaction, and global epistasis seen as the result of weak or no interactions (e.g., due to additive contributions to free energy) \cite{Sailer2018}.
Our work shows that, counterintuitively, global epistasis can also result from strong interactions, if the interactions result in a collective slow mode.

In conclusion, we have analyzed the evolutionary consequences of a  biophysical slow mode. Such slow modes are often linked to particular functions, such as allostery in proteins \cite{Wodak2019} or metabolic control of gene expression \cite{Erickson2017}. Our work suggests an alternative benefit of such slow modes, namely evolvability. This suggests that perhaps slow modes might have evolved for evolvability first, and only later been repurposed for other functions.

\indent \textbf{Acknowledgements} We are grateful to Rama Ranganathan, Sidney Nagel, Tobin Sosnick, Joseph Thornton, Weerapat Pittayakanchit, Yaakov Kleorin, and Lauren McGough, as well as the Ranganathan and Murugan groups, for helpful discussions, and to Tyler Starr and Jakub Otwinowski for feedback. KH thanks the James S. McDonnell Foundation for support via a postdoctoral fellowship. AM acknowledge support from the Simons Investigator program.

\bibliographystyle{unsrt}
\bibliography{epistasis.bib}

\begin{thebibliography}{10}

\bibitem{Starr2016}
Tyler~N. Starr and Joseph~W. Thornton.
\newblock {Epistasis in protein evolution}.
\newblock {\em Protein Science}, 25(7):1204--1218, jul 2016.

\bibitem{Adams2019}
Rhys~M. Adams, Justin~B. Kinney, Aleksandra~M. Walczak, and Thierry Mora.
\newblock {Epistasis in a Fitness Landscape Defined by Antibody-Antigen Binding
  Free Energy}.
\newblock {\em Cell Systems}, 8(1):86--93.e3, jan 2019.

\bibitem{Domingo2019}
J{\'{u}}lia Domingo, Pablo Baeza-Centurion, and Ben Lehner.
\newblock {The Causes and Consequences of Genetic Interactions (Epistasis)}.
\newblock {\em Annual Review of Genomics and Human Genetics},
  20(1):annurev--genom--083118--014857, aug 2019.

\bibitem{Macia2012}
Javier Mac{\'{i}}a, Ricard~V. Sol{\'{e}}, and Santiago~F. Elena.
\newblock {The causes of epistasis in genetic networks}.
\newblock {\em Evolution}, 66(2):586--596, feb 2012.

\bibitem{RojasEchenique2019}
Jos{\'{e}}~I. {Rojas Echenique}, Sergey Kryazhimskiy, Alex~N. {Nguyen Ba}, and
  Michael~M. Desai.
\newblock {Modular epistasis and the compensatory evolution of gene deletion
  mutants}.
\newblock {\em PLoS Genetics}, 15(2):e1007958, feb 2019.

\bibitem{Segre2005}
Daniel Segr{\`{e}}, Alexander DeLuna, George~M. Church, and Roy Kishony.
\newblock {Modular epistasis in yeast metabolism}.
\newblock {\em Nature Genetics}, 37(1):77--83, jan 2005.

\bibitem{Bajic2018}
Djordje Baji{\'{c}}, Jean~C.C. Vila, Zachary~D Blount, and Alvaro
  S{\'{a}}nchez.
\newblock {On the deformability of an empirical fitness landscape by microbial
  evolution}.
\newblock {\em Proceedings of the National Academy of Sciences of the United
  States of America}, 115(44):11286--11291, oct 2018.

\bibitem{Gould2018}
Alison~L Gould, Vivian Zhang, Lisa Lamberti, Eric~W Jones, Benjamin Obadia,
  Nikolaos Korasidis, Alex Gavryushkin, Jean~M Carlson, Niko Beerenwinkel, and
  William~B Ludington.
\newblock {Microbiome interactions shape host fitness.}
\newblock {\em Proceedings of the National Academy of Sciences of the United
  States of America}, 115(51):E11951--E11960, dec 2018.

\bibitem{Grilli2017}
Jacopo Grilli, Gy{\"{o}}rgy Barab{\'{a}}s, Matthew~J. Michalska-Smith, and
  Stefano Allesina.
\newblock {Higher-order interactions stabilize dynamics in competitive network
  models}.
\newblock {\em Nature}, 548(7666):210--213, aug 2017.

\bibitem{Kauffman1989}
Stuart~A. Kauffman and Edward~D. Weinberger.
\newblock {The NK model of rugged fitness landscapes and its application to
  maturation of the immune response}.
\newblock {\em Journal of Theoretical Biology}, 141(2):211--245, 1989.

\bibitem{Starr2017}
Tyler~N. Starr, Lora~K. Picton, and Joseph~W. Thornton.
\newblock {Alternative evolutionary histories in the sequence space of an
  ancient protein}.
\newblock {\em Nature}, 549(7672):409--413, sep 2017.

\bibitem{Starr2018}
Tyler~N. Starr, Julia~M. Flynn, Parul Mishra, Daniel~N.A. Bolon, and Joseph~W.
  Thornton.
\newblock {Pervasive contingency and entrenchment in a billion years of Hsp90
  evolution}.
\newblock {\em Proceedings of the National Academy of Sciences of the United
  States of America}, 115(17):4453--4458, apr 2018.

\bibitem{Otwinowski2018a}
Jakub Otwinowski, David~M. McCandlish, and Joshua~B. Plotkin.
\newblock {Inferring the shape of global epistasis}.
\newblock {\em Proceedings of the National Academy of Sciences of the United
  States of America}, 115(32):E7550--E7558, 2018.

\bibitem{Otwinowski2018}
Jakub Otwinowski.
\newblock {Biophysical inference of epistasis and the effects of mutations on
  protein stability and function}.
\newblock {\em Molecular Biology and Evolution}, 35(10):2345--2354, aug 2018.

\bibitem{Sailer2018}
Zachary~R Sailer and Michael~J Harms.
\newblock {Uninterpretable interactions: epistasis as uncertainty}.
\newblock {\em bioRxiv}, page 378489, 2018.

\bibitem{Kryazhimskiy2014}
Sergey Kryazhimskiy, Daniel~P. Rice, Elizabeth~R. Jerison, and Michael~M.
  Desai.
\newblock {Global epistasis makes adaptation predictable despite sequence-level
  stochasticity}.
\newblock {\em Science}, 344(6191):1519--1522, jun 2014.

\bibitem{Lehner2011}
Ben Lehner.
\newblock {Molecular mechanisms of epistasis within and between genes}.
\newblock {\em Trends in Genetics}, 27(8):323--331, aug 2011.

\bibitem{DePristo2005}
Mark~A. DePristo, Daniel~M. Weinreich, and Daniel~L. Hartl.
\newblock {Missense meanderings in sequence space: A biophysical view of
  protein evolution}.
\newblock {\em Nature Reviews Genetics}, 6(9):678--687, sep 2005.

\bibitem{Sailer2017b}
Zachary~R. Sailer and Michael~J. Harms.
\newblock {Molecular ensembles make evolution unpredictable}.
\newblock {\em Proceedings of the National Academy of Sciences of the United
  States of America}, 114(45):11938--11943, 2017.

\bibitem{Yeh2006}
Pamela Yeh, Ariane~I Tschumi, and Roy Kishony.
\newblock {Functional classification of drugs by properties of their pairwise
  interactions}.
\newblock {\em Nature Genetics}, 38(4):489--494, apr 2006.

\bibitem{Chait2007}
Remy Chait, Allison Craney, and Roy Kishony.
\newblock {Antibiotic interactions that select against resistance}.
\newblock {\em Nature}, 446(7136):668--671, apr 2007.

\bibitem{Wood2012}
Kevin Wood, Satoshi Nishida, Eduardo~D. Sontag, and Philippe Cluzel.
\newblock {Mechanism-independent method for predicting response to multidrug
  combinations in bacteria}.
\newblock {\em Proceedings of the National Academy of Sciences of the United
  States of America}, 109(30):12254--12259, jul 2012.

\bibitem{Sanchez-Gorostiaga2018}
Alicia Sanchez-Gorostiaga, Djordje Baji{\'{c}}, Melisa~L. Osborne, Juan~F
  Poyatos, and Alvaro Sanchez.
\newblock {High-order interactions dominate the functional landscape of
  microbial consortia}.
\newblock {\em bioRxiv}, page 333534, may 2018.

\bibitem{Bahar2010}
Ivet Bahar, Timothy~R. Lezon, Ahmet Bakan, and Indira~H. Shrivastava.
\newblock {Normal mode analysis of biomolecular structures: Functional
  mechanisms of membrane proteins}.
\newblock {\em Chemical Reviews}, 110(3):1463--1497, 2010.

\bibitem{Brooks1983}
B.~Brooks and M.~Karplus.
\newblock {Harmonic dynamics of proteins: Normal modes and fluctuations in
  bovine pancreatic trypsin inhibitor}.
\newblock {\em Proceedings of the National Academy of Sciences of the United
  States of America}, 80(21 I):6571--6575, nov 1983.

\bibitem{Hekstra2016}
Doeke~R. Hekstra, K.~Ian White, Michael~A. Socolich, Robert~W. Henning, Vukica
  {\v{S}}rajer, and Rama Ranganathan.
\newblock {Electric-field-stimulated protein mechanics}.
\newblock {\em Nature}, 540(7633):400--405, dec 2016.

\bibitem{Rocks2017}
Jason~W. Rocks, Nidhi Pashine, Irmgard Bischofberger, Carl~P. Goodrich,
  Andrea~J. Liu, and Sidney~R. Nagel.
\newblock {Designing allostery-inspired response in mechanical networks}.
\newblock {\em Proceedings of the National Academy of Sciences of the United
  States of America}, 114(10):2520--2525, mar 2017.

\bibitem{Dutta2018}
Sandipan Dutta, Jean~Pierre Eckmann, Albert Libchaber, and Tsvi Tlusty.
\newblock {Green function of correlated genes in a minimal mechanical model of
  protein evolution}.
\newblock {\em Proceedings of the National Academy of Sciences of the United
  States of America}, 115(20):E4559--E4568, 2018.

\bibitem{Wodak2019}
Shoshana~J. Wodak, Emanuele Paci, Nikolay~V. Dokholyan, Igor~N. Berezovsky,
  Amnon Horovitz, Jing Li, Vincent~J. Hilser, Ivet Bahar, John Karanicolas,
  Gerhard Stock, Peter Hamm, Roland~H. Stote, Jerome Eberhardt, Yassmine
  Chebaro, Annick Dejaegere, Marco Cecchini, Jean~Pierre Changeux, Peter~G.
  Bolhuis, Jocelyne Vreede, Pietro Faccioli, Simone Orioli, Riccardo Ravasio,
  Le~Yan, Carolina Brito, Matthieu Wyart, Paraskevi Gkeka, Ivan Rivalta, Giulia
  Palermo, J.~Andrew McCammon, Joanna Panecka-Hofman, Rebecca~C. Wade,
  Antonella {Di Pizio}, Masha~Y. Niv, Ruth Nussinov, Chung~Jung Tsai, Hyunbum
  Jang, Dzmitry Padhorny, Dima Kozakov, and Tom McLeish.
\newblock {Allostery in Its Many Disguises: From Theory to Applications}.
\newblock {\em Structure}, 27(4):566--578, apr 2019.

\bibitem{Yan2017}
Le~Yan, Riccardo Ravasio, Carolina Brito, and Matthieu Wyart.
\newblock {Architecture and coevolution of allosteric materials}.
\newblock {\em Proceedings of the National Academy of Sciences of the United
  States of America}, 114(10):2526--2531, mar 2017.

\bibitem{Rivoire2019}
Olivier Rivoire.
\newblock {Parsimonious evolutionary scenario for the origin of allostery and
  coevolution patterns in proteins}.
\newblock {\em Physical Review E}, 100(3):032411, sep 2019.

\bibitem{Leo-Macias2005}
Alejandra Leo-Macias, Pedro Lopez-Romero, Dmitry Lupyan, Daniel Zerbino, and
  Angel~R. Ortiz.
\newblock {An analysis of core deformations in protein superfamilies}.
\newblock {\em Biophysical Journal}, 88(2):1291--1299, feb 2005.

\bibitem{Echave2010}
Juli{\'{a}}n Echave and Francisco~M. Fern{\'{a}}ndez.
\newblock {A perturbative view of protein structural variation}.
\newblock {\em Proteins: Structure, Function and Bioinformatics},
  78(1):173--180, jan 2010.

\bibitem{Raimondi2010}
Francesco Raimondi, Modesto Orozco, and Francesca Fanelli.
\newblock {Deciphering the Deformation Modes Associated with Function Retention
  and Specialization in Members of the Ras Superfamily}.
\newblock {\em Structure}, 18(3):402--414, mar 2010.

\bibitem{Rod2003}
Thomas~H. Rod, Jennifer~L. Radkiewicz, and Charles~L. Brooks.
\newblock {Correlated motion and the effect of distal mutations in
  dihydrofolate reductase}.
\newblock {\em Proceedings of the National Academy of Sciences of the United
  States of America}, 100(12):6980--6985, jun 2003.

\bibitem{Yang2008}
Lei Yang, Guang Song, Alicia Carriquiry, and Robert~L. Jernigan.
\newblock {Close Correspondence between the Motions from Principal Component
  Analysis of Multiple HIV-1 Protease Structures and Elastic Network Modes}.
\newblock {\em Structure}, 16(2):321--330, feb 2008.

\bibitem{Sinha2002}
Neeti Sinha and Ruth Nussinov.
\newblock {Point mutations and sequence variability in proteins:
  Redistributions of preexisting populations}.
\newblock {\em Proceedings of the National Academy of Sciences of the United
  States of America}, 98(6):3139--3144, mar 2001.

\bibitem{Eckmann2019}
Jean-Pierre Eckmann, Jacques Rougemont, and Tsvi Tlusty.
\newblock { Colloquium : Proteins: The physics of amorphous evolving matter }.
\newblock {\em Reviews of Modern Physics}, 91(3):031001, jul 2019.

\bibitem{Tobi2005}
Dror Tobi and Ivet Bahar.
\newblock {Structural changes involved in protein binding correlate with
  intrinsic motions of proteins in the unbound state}.
\newblock {\em Proceedings of the National Academy of Sciences of the United
  States of America}, 102(52):18908--18913, 2005.

\bibitem{Bakan2009}
Ahmet Bakan and Ivet Bahar.
\newblock {The intrinsic dynamics of enzymes plays a dominant role in
  determining the structural changes induced upon inhibitor binding}.
\newblock {\em Proceedings of the National Academy of Sciences of the United
  States of America}, 106(34):14349--14354, 2009.

\bibitem{Wang2019}
Shou-Wen Wang, Anne-Florence Bitbol, and Ned~S. Wingreen.
\newblock {Revealing evolutionary constraints on proteins through sequence
  analysis}.
\newblock {\em PLOS Computational Biology}, 15(4):e1007010, apr 2019.

\bibitem{Tirion1996}
Monique~M. Tirion.
\newblock {Large amplitude elastic motions in proteins from a single-parameter,
  atomic analysis}.
\newblock {\em Physical Review Letters}, 77(9):1905--1908, aug 1996.

\bibitem{Best2006}
Robert~B. Best, Kresten Lindorff-Larsen, Mark~A. DePristo, and Michele
  Vendruscolo.
\newblock {Relation between native ensembles and experimental structures of
  proteins}.
\newblock {\em Proceedings of the National Academy of Sciences of the United
  States of America}, 103(29):10901--10906, jul 2006.

\bibitem{Schmidt2016}
Alexander Schmidt, Karl Kochanowski, Silke Vedelaar, Erik Ahrn{\'{e}}, Benjamin
  Volkmer, Luciano Callipo, K{\`{e}}vin Knoops, Manuel Bauer, Ruedi Aebersold,
  and Matthias Heinemann.
\newblock {The quantitative and condition-dependent Escherichia coli proteome}.
\newblock {\em Nature Biotechnology}, 34(1):104--110, jan 2016.

\bibitem{Barenholz2016}
Uri Barenholz, Leeat Keren, Eran Segal, and Ron Milo.
\newblock {A minimalistic resource allocation model to explain ubiquitous
  increase in protein expression with growth rate}.
\newblock {\em PLoS ONE}, 11(4):e0153344, apr 2016.

\bibitem{Kaneko2015}
Kunihiko Kaneko, Chikara Furusawa, and Tetsuya Yomo.
\newblock {Universal relationship in gene-expression changes for cells in
  steady-growth state}.
\newblock {\em Physical Review X}, 5(1), 2015.

\bibitem{Furusawa2018}
Chikara Furusawa and Kunihiko Kaneko.
\newblock {Formation of dominant mode by evolution in biological systems}.
\newblock {\em Physical Review E}, 97(4):42410, 2018.

\bibitem{Kaneko2018}
Kunihiko Kaneko and Chikara Furusawa.
\newblock {Macroscopic Theory for Evolving Biological Systems Akin to
  Thermodynamics}.
\newblock {\em Annual Review of Biophysics}, 47(1):273--290, may 2018.

\bibitem{Sato2019}
Takuya~U. Sato and Kunihiko Kaneko.
\newblock {Evolutionary dimension reduction in phenotypic space}.
\newblock 2019.

\bibitem{You2013}
Conghui You, Hiroyuki Okano, Sheng Hui, Zhongge Zhang, Minsu Kim, Carl~W.
  Gunderson, Yi-Ping Wang, Peter Lenz, Dalai Yan, and Terence Hwa.
\newblock {Coordination of bacterial proteome with metabolism by cyclic AMP
  signalling}.
\newblock {\em Nature}, 500(7462):301--306, aug 2013.

\bibitem{Erickson2017}
David~W. Erickson, Severin~J. Schink, Vadim Patsalo, James~R. Williamson,
  Ulrich Gerland, and Terence Hwa.
\newblock {A global resource allocation strategy governs growth transition
  kinetics of Escherichia coli}.
\newblock {\em Nature}, 551(7678):119--123, nov 2017.

\bibitem{Togashi2007}
Yuichi Togashi and Alexander~S Mikhailov.
\newblock {Nonlinear relaxation dynamics in elastic networks and design
  principles of molecular machines.}
\newblock {\em Proceedings of the National Academy of Sciences of the United
  States of America}, 104(21):8697--702, may 2007.

\bibitem{Olson2014}
C.~Anders Olson, Nicholas~C. Wu, and Ren Sun.
\newblock {A comprehensive biophysical description of pairwise epistasis
  throughout an entire protein domain}.
\newblock {\em Current Biology}, 24(22):2643--2651, nov 2014.

\bibitem{Salinas2018}
Victor~H Salinas and Rama Ranganathan.
\newblock {Coevolution-based inference of amino acid interactions underlying
  protein function}.
\newblock {\em eLife}, 7, jul 2018.

\bibitem{Diss2018}
Guillaume Diss and Ben Lehner.
\newblock {The genetic landscape of a physical interaction}.
\newblock {\em eLife}, 7, 2018.

\bibitem{Chure2019}
Griffin Chure, Manuel Razo-Mejia, Nathan~M. Belliveau, Tal Einav, Zofii~A.
  Kaczmarek, Stephanie~L. Barnes, Mitchell Lewis, and Rob Phillips.
\newblock {Predictive shifts in free energy couple mutations to their
  phenotypic consequences}.
\newblock {\em Proceedings of the National Academy of Sciences},
  116(37):18275--18284, sep 2019.

\end{thebibliography}


\begin{thebibliography}{10}

\bibitem{Otwinowski2018a}
Jakub Otwinowski, David~M. McCandlish, and Joshua~B. Plotkin.
\newblock {Inferring the shape of global epistasis}.
\newblock {\em Proceedings of the National Academy of Sciences of the United
  States of America}, 115(32):E7550--E7558, 2018.

\bibitem{Dutta2018}
Sandipan Dutta, Jean~Pierre Eckmann, Albert Libchaber, and Tsvi Tlusty.
\newblock {Green function of correlated genes in a minimal mechanical model of
  protein evolution}.
\newblock {\em Proceedings of the National Academy of Sciences of the United
  States of America}, 115(20):E4559--E4568, 2018.

\bibitem{Yan2017}
Le~Yan, Riccardo Ravasio, Carolina Brito, and Matthieu Wyart.
\newblock {Architecture and coevolution of allosteric materials}.
\newblock {\em Proceedings of the National Academy of Sciences of the United
  States of America}, 114(10):2526--2531, mar 2017.

\bibitem{Rocks2017}
Jason~W. Rocks, Nidhi Pashine, Irmgard Bischofberger, Carl~P. Goodrich,
  Andrea~J. Liu, and Sidney~R. Nagel.
\newblock {Designing allostery-inspired response in mechanical networks}.
\newblock {\em Proceedings of the National Academy of Sciences of the United
  States of America}, 114(10):2520--2525, mar 2017.

\bibitem{Bahar2010}
Ivet Bahar, Timothy~R. Lezon, Ahmet Bakan, and Indira~H. Shrivastava.
\newblock {Normal mode analysis of biomolecular structures: Functional
  mechanisms of membrane proteins}.
\newblock {\em Chemical Reviews}, 110(3):1463--1497, 2010.

\bibitem{Leo-Macias2005}
Alejandra Leo-Macias, Pedro Lopez-Romero, Dmitry Lupyan, Daniel Zerbino, and
  Angel~R. Ortiz.
\newblock {An analysis of core deformations in protein superfamilies}.
\newblock {\em Biophysical Journal}, 88(2):1291--1299, feb 2005.

\bibitem{Kaneko2015}
Kunihiko Kaneko, Chikara Furusawa, and Tetsuya Yomo.
\newblock {Universal relationship in gene-expression changes for cells in
  steady-growth state}.
\newblock {\em Physical Review X}, 5(1), 2015.

\bibitem{Furusawa2018}
Chikara Furusawa and Kunihiko Kaneko.
\newblock {Formation of dominant mode by evolution in biological systems}.
\newblock {\em Physical Review E}, 97(4):42410, 2018.

\bibitem{Kaneko2018}
Kunihiko Kaneko and Chikara Furusawa.
\newblock {Macroscopic Theory for Evolving Biological Systems Akin to
  Thermodynamics}.
\newblock {\em Annual Review of Biophysics}, 47(1):273--290, may 2018.

\bibitem{Segre2005}
Daniel Segr{\`{e}}, Alexander DeLuna, George~M. Church, and Roy Kishony.
\newblock {Modular epistasis in yeast metabolism}.
\newblock {\em Nature Genetics}, 37(1):77--83, jan 2005.

\bibitem{Chure2019}
Griffin Chure, Manuel Razo-Mejia, Nathan~M. Belliveau, Tal Einav, Zofii~A.
  Kaczmarek, Stephanie~L. Barnes, Mitchell Lewis, and Rob Phillips.
\newblock {Predictive shifts in free energy couple mutations to their
  phenotypic consequences}.
\newblock {\em Proceedings of the National Academy of Sciences},
  116(37):18275--18284, sep 2019.

\bibitem{Schmidt2016}
Alexander Schmidt, Karl Kochanowski, Silke Vedelaar, Erik Ahrn{\'{e}}, Benjamin
  Volkmer, Luciano Callipo, K{\`{e}}vin Knoops, Manuel Bauer, Ruedi Aebersold,
  and Matthias Heinemann.
\newblock {The quantitative and condition-dependent Escherichia coli proteome}.
\newblock {\em Nature Biotechnology}, 34(1):104--110, jan 2016.

\bibitem{Barenholz2016}
Uri Barenholz, Leeat Keren, Eran Segal, and Ron Milo.
\newblock {A minimalistic resource allocation model to explain ubiquitous
  increase in protein expression with growth rate}.
\newblock {\em PLoS ONE}, 11(4):e0153344, apr 2016.

\end{thebibliography}

\end{document}



\title{ {\Large Physical constraints on epistasis} \\ {\normalsize SI} }

\author{K. Husain, A. Murugan}
\affiliation{Department of Physics, the University of Chicago}

\maketitle 

\tableofcontents

\section{Global and low-rank epistasis in proteins with a soft mechanical mode}

\subsection{Theory}
\indent We begin, as in the main text, with proteins. Our goal is to relate the epistatic co-efficient, $\Delta\Delta F_{ij}$, of some (arbitrary) function (e.g. binding affinity for a ligand) to the mechanical modes of the protein. Recall the sequence-to-function map of the protein
\be
\text{ sequence } \mbf{s} \to \text{ function } F \l \mbf{s} \r
\ee
from which we define epistasis in the usual single reference manner, as an expansion of the fitness function $F(\mbf{s})$ around the wildtype genotype (here, $\mbf{s}$ is a binary vector representation of the genotype, where $s_i = 1$ if residue $i$ is mutated and $0$ otherwise):
\be
F(\mbf{s}) \equiv F(\mbf{r}(\mbf{s})) = \underbrace{F_0}_{\text{WT fitness}} + \sum_i \Delta F_i \, s_i + \sum_{i > j} \Delta\Delta F_{ij} \, s_is_j + \sum_{i>j>k} \Delta\Delta\Delta F_{ijk} \, s_is_js_k + \ldots
\ee 
\indent Key to the calculation is to express the function of the protein (e.g., its binding affinity or enzymatic activity), $F$, in terms of its physical structure, schematically denoted $\mathbf{r}$ (we shall more precisely define $\mathbf{r}$ in a moment): $F(\mathbf{r})$. The sequence-to-function map then conceptually decomposes into two steps: sequence to structure, and structure to function:
\be
\text{sequence } \mathbf{s} \to \text{ structure } \mathbf{r}(\mbf{s})  \to \text{ function } F \l \mbf{r}\l \mbf{s} \r \r \nn
\ee
\indent Before proceeding, we note that the function of the protein may additionally depend on sequence directly, i.e. not through the structure $\mbf{r}$, $F(\mbf{r}(\mbf{s}), \mbf{s})$. For instance, the precise chemical identity of the amino acid at a particular site may be crucial -- however, it is difficult to imagine that this is true outside of the small number of residues that form the protein's active site. Thus, for the remainder of the paper, we neglect this potential direct dependence.

\subsubsection*{Mutation-induced deformations}

First, we compute the structural deformation induced by mutations in a protein, in a small perturbation approximation. Represent the structure of the protein by a single vector $\mathbf{r}$ -- containing, say, the concatenated spatial positions of all $N$ atoms in the structure. In the absence of external forcing, the structure relaxes in the folding energy $E \l \mathbf{r} \r$:
\be \label{eq:justGradientDescent}
\f{\p}{\p t}\mathbf{r} = - \grad E \l \mathbf{r} \r
\ee
\indent The sequence specifies the form of the energy $E$ -- we shall discuss this dependence in a moment, but first let us denote the wildtype energy by $E_0$. The wildtype structure $\mathbf{r}_0$ is found at the minimum of $E_0(\mathbf{r})$, satisfying:
\be
-\grad E_0 \l \mathbf{r}_0 \r = 0
\ee
\indent Consider now a mutation (or multiple mutations), which perturbs the folding energy function $E(\mbf{r})$. We compute the resultant deformation in structure. This is easily done if we use a book-keeping parameter $\epsilon$. Denote the modified energy function by $E\l \mathbf{r} \r = E_0\l \mathbf{r} \r + \epsilon\delta E \l \mathbf{r} \r$. We look for a new minimum perturbatively in $\epsilon$: $\mathbf{r} = \mathbf{r}_0 + \epsilon \delta \mathbf{r} + \ldots$. Expanding Eq. \ref{eq:justGradientDescent} to first order in $\epsilon$ at steady state:
\bea
0 &=& -\grad E_0 \l \mbf{r}_0 + \epsilon\delta\mbf{r} + \ldots \r - \epsilon\grad \delta E \l \mbf{r}_0 + \epsilon\delta\mbf{r} + \ldots \r \nn \\
&\approx& -\grad E_0 \l \mathbf{r}_0 \r - \epsilon \l \mathcal{H}_0 \, \delta\mbf{r} + \grad\delta E \l \mbf{r}_0 \r \r
\eea
\noindent where the matrix $\mathcal{H}_0$ is the Hessian of the wildtype energy $E_0 \l \mbf{r} \r$ evaluated at the wildtype structure $\mbf{r}_0$. The first term is identically $0$, and so we obtain for the mutation-induced deformation:
\be \label{eq:generalDeformation}
\delta\mbf{r} = - \mathcal{H}_0^{-1} \grad \delta E \l \mbf{r}_0 \r
\ee
\noindent where the inverse $\mathcal{H}_0^{-1}$ is to be understood in the Morse-Penrose sense (i.e. projecting out the trivial zero-modes corresponding to global rotations and translations).

\indent \textbf{Now let us suppose that the wildtype protein has a single soft mode.} Mathematically, this corresponds to an eigenvector of $\mathcal{H}_0$, $\hat{v}_0$, with an eigenvalue $\lambda_0$ much smaller than any other eigenvalue $\lambda_m$, $m \neq 0$. Expanding $\mathcal{H}_0^{-1}$ in the eigenbasis and keeping only the largest term (which corresponds to the eigenmode $\hat{v}_0$):
\bea \label{eq:deformingPhi}
\delta \mbf{r} &=& -\l \f{1}{\lambda_0} \hat{v}_0 \hat{v}_0^T + \cdots \r \grad \delta E \nn \\
 &=& -\sum_{m = 0}^{3N-7} \f{1}{\lambda_m} \hat{v}_m \, \l \hat{v}_m \cdot \grad\delta E \r \nn \\
&\approx& \f{1}{\lambda_0} \hat{v}_0 \underbrace{\l - \hat{v}_0 \cdot \grad \delta E \r}_{\phi}
\eea
\noindent where we have defined the deformation magnitude along the soft mode, $\phi$. Thus, as advertised in the main text, mutations in the background of the soft mode deform the protein preferentially along that mode, with a magnitude given by $\phi$. 

\subsubsection*{Global epistasis}

\indent To proceed, we need to discuss how the deformation magnitude $\phi$ depends on the number of mutations made. For concreteness, let us introduce the binary variable $s_i = 0$ ($1$) when residue $i$ is wildtype (mutant). The vector $\mbf{s}$, with elements $s_i$, thus represents the sequence of the protein. If we suppose that the energy function $E(\mbf{r})$ is a sum of pairwise terms between atoms, then the most general form for the perturbed energy function $\delta E$ is
\be \label{eq:mutAndEnergy}
\delta E = \sum_i \delta_i E \, s_i + \sum_{i > j} \delta_{ij} E \, s_i s_j
\ee
\noindent where $\delta_i E$ is the change in energy due to the mutation $i$ alone, and $\delta_{ij} E$ is the contribution of the double mutant $i$ and $j$. Notably, $\delta_{ij} E$ will only be non-zero only when residues $i$ and $j$ are directly interacting physically. As we suppose that physical interactions are local, $\delta_{ij} E$ is thus non-zero only when $i$ and $j$ are in contact -- that is, the matrix $\delta_{ij} E$ is sparse and has the structure of the contact matrix of the protein. 

\indent Inserting the expression for $\delta E$ into Eq. \ref{eq:deformingPhi}:
\be \label{eq:deformationMutations}
\phi = \sum_i \underbrace{-\hat{v}_0 \cdot \grad \delta_i E}_{\theta_i} \, s_i + \sum_{i > j} \underbrace{-\hat{v}_0 \cdot \grad \delta_{ij} E}_{C_{ij}} \, s_i s_j
\ee
\noindent which serves as a defintion of $\theta_i$ and $C_{ij}$. Note that the deformation magnitude $\phi$ is nearly additive in mutations -- differing only by a sparse second-order term induced by physical contacts.

\indent Now consider a fitness function for the protein (representing, for instance, the binding affinity of the protein for a ligand, or the catalytic rate of an enzymatic reaction). In general, this quantitative trait is a function of both protein structure and sequence: $F\l \mathbf{r}, \mathbf{s} \r$. Let us assume, however, the direct dependence on sequence is negligible: likely involving only residues directly at the active site of the protein. We thus write it as only a function of structure $F \l \mbf{r} \r$.

\indent The fitness of a mutated sequence, with sequence $\mbf{s}$ and structure $\mbf{r}_0 + \delta \mbf{r}$ is $F(\mbf{r}_0 + \delta \mbf{r}(\mbf{s}))$. We use Eqs. \ref{eq:deformingPhi} and \ref{eq:deformationMutations} to rewrite the fitness in terms of the scalar variable $\phi$:
\bea \label{eq:globalEpistasis}
&& g\l \phi \r \equiv F\l \mbf{r}_0 + \f{\hat{v}_0}{\lambda_0} \phi \r \nn \\
&& \phi \l \mbf{s} \r = \sum_i \theta_i \, s_i + \sum_{i > j} C_{ij} \, s_i s_j
\eea
\noindent which, except for the sparse matrix $C_{ij}$, is in the standard form for global epistasis, as defined in, e.g., \cite{Otwinowski2018a}.

\subsubsection*{Low-rank epistasis}

\indent Let us further suppose that $\phi$ is small and we can Taylor expand $g(\phi)$ to second order. Grouping terms by the order of mutations (i.e. $s_i$):
\be
g(\phi) \approx g(0) + \sum_i \l g^{\prime}(0) + g^{\prime\prime} (0)  \, \theta_i \r \,\theta_i\, s_i + \sum_{i > j} g^{\prime \prime}(0) \l \theta_i \theta_j + c_{ij} \r \, s_i s_j
\ee
\noindent where $c_{ij}$ are terms that involve $C_{ij}$ and correspondingly have the structure of the contact matrix. 

\indent By comparison with the standard form of an epistatic expansion:
\be
F(s_i) = F_0 + \sum \Delta F_i \, s_i + \sum \Delta \Delta F_{ij} \, s_i s_j + \cdots
\ee
\noindent we arrive at the low-rank form of epistasis advertised in the main text:
\be \label{eq:lowrankEpi}
\Delta \Delta F_{ij} \propto \theta_i\theta_j + c_{ij}
\ee
\indent A slightly different derivation of Eq. \ref{eq:lowrankEpi} is instructive. For simplicity, let us neglect the contact term in Eq. \ref{eq:mutAndEnergy}, writing the structural deformation as a function of sequence as:
\be
\delta \mbf{r} \l \mbf{s} \r = \sum_i \delta_i \mbf{r} \, s_i 
\ee
\indent Then,
\be
F \l \mbf{r}_0 + \delta \mbf{r} \r \approx F(\mbf{r}_0) + \sum_i \l \delta\mbf{r}_i\cdot\grad \r F(\mbf{r}_0) \, s_i + \f{1}{2}\sum_{i,j} \l \delta\mbf{r}_i\cdot\grad \r \l \delta\mbf{r}_j\cdot\grad \r F(\mbf{r}_0) \, s_i s_j
\ee
\indent Notice that we have not yet invoked the existence of a soft mode. In braket notation, the co-efficient of $s_is_j$ -- the epistatic co-efficient -- is $\braket{\delta_i \mbf{r} | \grad^2 F | \delta_j \mbf{r}}$, where the matrix $\grad^2 F$ is the Hessian of the fitness function, evaluated at $\mbf{r}_0$. Expanding $\delta_i \mbf{r}$ in terms of the WT modes $\hat{v}_m$:
\bea
\Delta\Delta F_{ij} &\sim& \braket{\delta_i \mbf{r} | \grad^2 F | \delta_j \mbf{r}} \nn \\
&=& \sum_{m = 0}^{3N - 7} \f{1}{\lambda_m^2} \, \braket{\hat{v}_m | \grad^2 F | \hat{v}_m}  \, \braket{\hat{v}_m | \grad\delta_iE} \braket{\hat{v}_m | \grad\delta_j E } \nn \\
&\approx& \f{1}{\lambda_0^2} \, \braket{\hat{v}_0 | \grad^2 F | \hat{v}_0} \, \theta_i \theta_j + \mathcal{O}\l \f{1}{\lambda_1^2} \r
\eea
\noindent where the final simplification holds only when there is a sufficiently (see next section) soft mode. The expression above makes clear that without a soft mode, each $\Delta \Delta F_{ij}$ is a different matrix element of the matrix $\grad^2 F$, and is thus an arbitrary number that depends on the idiosyncratic deformations $\delta_i \mbf{r}$, $\delta_j \mbf{r}$. With a soft mode, however, the matrix element reduces to a const. $\times$ the product of scalars, $\theta_i \theta_j$ -- and thus the epistatic matrix $\Delta \Delta F_{ij}$ is rank 1.

\subsubsection*{How big a gap?}

\indent How large a gap in the normal mode spectrum is required to obtain global epistasis? Consider Eq. \ref{eq:deformingPhi},
\be
\delta \mbf{r} = \f{1}{\lambda_0} \hat{v}_0 \, \l \hat{v}_0 \cdot \mbf{g} \r + \sum_{m = 1}^{M-1} \f{1}{\lambda_m} \hat{v}_m \, \l \hat{v}_m \cdot \mbf{g} \r
\ee
\noindent where we have defined the mutation-induced force $\mbf{g}\equiv -\grad \delta E$, and use $M$ to represent the number of modes ($M = 3N-6$ for $N$ atoms in 3D).

\indent As argued above, global epistasis arises when the deformation $\delta \mbf{r}$ is along the soft mode $\hat{v}_0$ for an arbitrary mutation-induced force $\mbf{g}$. That is, the projection of $\delta \mbf{r}$ onto $\hat{v}_0$ is of much greater magnitude than the projection orthogonal to $\hat{v}_0$:
\be
\f{1}{\lambda_0^2} \l \hat{v}_0 \cdot \mbf{g} \r^2 \gg \sum_{m=1}^{M-1} \f{1}{\lambda_m^2} \l \hat{v}_m \cdot \mbf{g} \r^2
\ee
\indent We compute the required mode-gap in the following manner. Let us suppose that all $\lambda_m$, $m > 0$, are of similar magnitude $\lambda_1 > \lambda_0$. As we are interested in a typical mutation-induced force $\mbf{g}$, we estimate $\hat{v}_m \cdot \mbf{g}$ by the average magnitude of two random vectors in a space of dimension $dN$ (where $d$ is the spatial dimension and $N$ is the number of atoms), denoted by $\alpha$:
\be
\f{1}{\lambda_0^2} \alpha^2 \gg \l M - 1 \r \f{1}{\lambda_1^2} \alpha^2
\ee
\indent Rearranging, we arrive at:
\be
\Delta E \equiv \f{\lambda_1}{\lambda_0} \gg \sqrt{M -1}
\ee
\indent As the number of modes $M$ grows linearly with the number of components of the system, $N$, we arrive at the $\sqrt{N}$ scaling advertised in the main text.

\subsection{Simulation}

\subsubsection*{Model}

\indent To mimic protein mechanics, we follow in the footsteps of recent work \cite{Dutta2018,Yan2017,Rocks2017} and simulate a 2D arrangement of nodes and springs (a so-called `elastic network model' or ENM), where nodes represent, e.g., the C-$\alpha$ atoms of each residue. Indexing nodes by $i$, the energy is:
\be\label{eq:springEnergy}
E = \f{1}{2} \sum_{i > j} k_{ij} \, \l \vert \mathbf{r}_i - \mathbf{r}_j \vert - l_{ij} \r ^2 
\ee
\noindent where the spring constant $k_{ij} = k$ if $i$, $j$ are in contact and $0$ otherwise, and $l_{ij}$ is the rest-length of the spring between $i$ and $j$. We choose $k = 1$, and set the rest length $l_{ij}$ by the distance between $i$ and $j$ in the wildtype structure $\mbf{r}_0$ (that is, the wildtype is unstrained).

\indent All spring networks are initialised with a slightly perturbed hexagonal lattice, of size $9\times 9$ nodes, with each node connected by springs to its nearest neighbours. All springs are initialised with rest lengths such that the network, in its resting configuration, is unstressed. Units of length are set by the average distance between a node and its nearest neighbour. We treat all deformations in the linear approximation, as is commonly done with elastic network models of protein mechanics \cite{Dutta2018,Yan2017,Bahar2010}.

\subsubsection*{Obtaining a soft mode: Monte Carlo algorithm}

\indent We engineer in a soft mode by a simple Monte Carlo algorithm. Briefly, at each step the algorithm selects a node and displaces it by a small amount $\delta \mbf{x}$, each of whose components is picked uniformly from the interval $\left[ -0.025, 0.025 \right]$. The rest lengths of all springs connected to the node are recomputed so that the structure remains unstressed. The Hessian of the network is computed and diagonalised, and the ratio of the two lowest modes is computed, 
\be
\Delta E \equiv \f{\lambda_1}{\lambda_0}
\ee
\noindent after excluding the 3 modes corresponding to global translations and rotations, that have eigenvalue $0$. The move is accepted if two criteria are met:

\begin{enumerate}
	\item $\Delta E$ increases
	\item The inverse participation ratio of the softest mode $\hat{v}_0$, defined as $\sum (v_0)_i^4/\sum (v_0)_i^2$, decreases, or is already below a set threshold (here taken to be $2/N$, where $N = 9\times 9 = 81$ is the number of nodes). This is to prevent the formation of a localised soft mode, as opposed to the desired collective motion.
\end{enumerate}

If these criteria are met, the move is accepted; otherwise, it is rejected. The process continues until the desired value of $\Delta E$ is achieved.

\subsubsection*{Mutations in the ENM}

\indent We simulate the effect of mutations in our ENM by first supposing that each residue has only one possible mutant -- a simplification of the more complex $21$ possible amino acids. The mutant residue at each site is either `larger' or `smaller' than the wildtype residue -- functionally, this corresponds to a choice of a number $\sigma_i = \pm \sigma$ for each residue, where the sign is randomly chosen. $\sigma_i$ represents the fractional change in amino acid size upon mutation of residue $i$.

\indent That is, for a given sequence $\mbf{s}$, each rest length $l_{ij}$ is modified from its wildtype value:

\be \label{eq:restLengthMutants}
l_{ij} \to \l 1 + \sigma_i \r^{s_j} \l 1 + \sigma_j \r^{s_j} \, l_{ij}
\ee

\indent We choose $\sigma = 0.1$, corresponding to a roughly $10 \%$ difference in size between the wildtype and mutant amino acids.

\indent To compute the deformation upon mutation, we update the rest lengths and then compute the force $\mbf{f} = -\grad E_{\text{mut}}(\mbf{r}_0)$ experienced by the mutant protein in the wildtype configuration $\mbf{r}_0$ (where $E_{mut}$ is Eq. \ref{eq:springEnergy}, with rest lengths given by Eq. \ref{eq:restLengthMutants}). Mathematically, this is identical to $-\grad \delta E \l \mbf{r}_0 \r$, as defined in \ref{eq:mutAndEnergy}. The mutation-induced deformation is then computed from Eq. \ref{eq:generalDeformation}. 

\subsubsection*{Fitness functions, double-mutant cycles, and analysis of epistatic matrix}

\indent To compute the epistatic matrix, we perform a complete set of double mutant cycles for each simulated elastic network, recording -- for each pair of mutants $i$ and $j$ -- the structural deformations induced by the single mutants, $\delta_i \mbf{r}$ and $\delta_j \mbf{r}$, as well as that elicited by the double mutant, $\delta \mbf{r}_{ij}$.

\indent We then measure the fitness of each (single or double) mutant by a weighted mean-squared displacement from the wildtype protein:
\be
F(\mbf{r}) \equiv F(\mbf{r}_0 + \delta \mbf{r}) = -\f{1}{N}\sum_{\alpha = 1}^{N} \kappa_{\alpha} \vert \delta \mbf{r}_{\alpha} \vert^2
\ee
\noindent where $N = 9\times 9 = 81$ is the number of residues, indexed by $\alpha$. By construction, the wildtype has fitness $0$. We choose each $\kappa_{\alpha}$ uniformly from the interval $\left[ 0, 1 \right]$. Note that the choice of fitness function made here is not essential for our results.

\indent The epistatic co-efficient $\Delta\Delta F_{ij}$ is then computed in the standard way:
\be
\Delta\Delta F_{ij} = F\l \mbf{r}_0 + \delta \mbf{r}_{ij} \r - \l F \l \mbf{r}_0 + \delta_i \mbf{r} \r + F \l \mbf{r}_0 + \delta_j \mbf{r} \r - F \l \mbf{r}_0 \r \r
\ee
\indent To give the numbers an appropriate scale, we normalise the epistatic matrix by the average magnitude of the first order mutational effects (i.e $\langle \vert \Delta F_i \vert \rangle$).

\indent To analyse the rank of the epistatic matrix, we first compute $\Delta \Delta F^{(1)}_{ij}$ -- a rank-1 approximation of $\Delta\Delta F_{ij}$ -- from the singular valued decomposition $\Delta\Delta F = U\,S\,V^{T}$, where $S$ is a diagonal matrix with entries $\{s_1, s_2,\ldots \}$, with $s_1 > s_2 >\ldots$, as:
\be
\Delta\Delta F^{(1)} = U \times \, \text{diag} \l s_1, 0, \ldots, 0 \r \times \, V^{T}
\ee
\indent The matrix $\Delta\Delta F^{(1)}$ is known to be the `best' rank-1 approximation of $\Delta\Delta F$ (i.e. minimises the Frobenius norm of $\Delta\Delta F^{(1)} - \Delta\Delta F$).

\indent We then assessed the complexity of epistasis by computing:
\be \label{eq:kappadef}
\kappa = \f{\text{median} \l \vert \Delta\Delta F_{ij} - \Delta\Delta F^{(1)}_{ij} \vert \r}{\langle \vert \Delta \Delta F_{ij} \vert \rangle}
\ee
\noindent which is plotted, for each fitness function and each simulated elastic network, in Fig. 2 of the main text.

\subsection{ENM-derived normal mode analysis of PDB structures}

\textit{Pertaining to Figure 2a of the main text}

\indent We compute the normal mode spectrum of a protein from an elastic network model (ENM) built from the PDB structure of the protein. We choose the simplest variant of a ENM; briefly, a unstrained spring with unit spring constant is placed between C$-\alpha$ atoms within $10$ angstroms of each other in the crystal structure. The Hessian of the resulting elastic network is computed and diagonalised to obtain the normal mode energies (i.e. the eigenvalues). 

\subsection{Relating evolutionary and physical deformations}
\textit{Pertaining to Figure 2b of the main text}

\indent In Fig. 2b we reproduce results from \cite{Leo-Macias2005}, which related the physical deformations of a protein family (as computed from an elastic network based normal mode analysis) to the structural deformations seen between members of a protein family. In particular, we replot:

\begin{itemize}
	\item \textbf{i} The number of principal components required to explain 70\% of the structural variation seen across a protein family (quantified from Table 1 in \cite{Leo-Macias2005}, shown in orange in Fig. 2b,i in the main text). As a comparison, we computed the same for a randomly generated dataset of 200 protein family with similar sampling as considered in \cite{Leo-Macias2005} (i.e. 25 protein structures per family, 100 residues per protein), in which the structural changes of each residue are drawn from a normal distribution with zero mean and unit variance. This is plotted in grey in Fig. 2b,i in the main text.
	\item \textbf{ii} The z-score of the relationship between the top principal components (as defined above) in a protein family, and the lowest normal modes. The plot in Fig. 2b,ii of the main text is obtained by digitising Fig. 6c from \cite{Leo-Macias2005}; see \cite{Leo-Macias2005} for details of analysis. 
\end{itemize}

\section{Global and low-rank epistasis in protein ensembles}

\subsection{Theory}

\indent Our results above take a mechanical viewpoint of protein function, with a single ground state structure whose low energy vibrational excitations (i.e. its soft mechanical modes) stereotype and simplify epistasis. In contrast, protein function can also be analysed from the point of view of an \textit{ensemble} of protein conformations. Labelling different, discrete conformations by $a$, $b$..., the protein is statistically described by a (normalised) ensemble $\psi_a$, which is the probability of finding the protein in state $a$. 

\indent The function $F$ of a protein may quantitatively be described as $F(\psi_a)$ -- that is, different occupancies of the conformational ensemble can change the functional ability of the protein. A classic example is of allostery in hemoglobin, in which allosteric communication from neighbouring hemoglobin molecules shifts an internal equilibrium to alter its affinity for oxygen. 

\indent Generally speaking, the distribution $\psi_a$ obeys a master equation of the form:
\be
    \frac{d\psi_a}{dt} = \sum_{b \neq a} M_{ba} \psi_b - \sum_{b \neq a} M_{ab} \psi_a
\ee
\noindent where $M_{ab}$ is the transition rate from state $a$ to state $b$. In accordance with detailed balance, the transition rates obey $M_{ab}/M_{ba} = \exp \l \beta \Delta E \r$, where $\beta$ is the inverse temperature and $\Delta E \equiv E_a - E_b$ is the difference in energies between states $a$ and $b$.

\indent By the conservation of probability, and assuming that the ensemble is ergodic, the matrix $M_{ab}$ has a single, unique eigenvector with zero eigenvalue. This eigenvector, $\psi^{\text{(WT)}}_a$, is the steady-state (Boltzmann) distribution, describing the native-state thermal ensemble. The remaining eigenvectors, with eigenvalues $\lambda_p$, $p \in \{1,...\}$, correspond to distinct modes of relaxation to the native-state ensemble, with characteristic rates $\lambda_p$. If $\vert \lambda_1 \vert \ll \vert \lambda_1\vert$, any perturbation of the steady state ensemble quickly relaxes to the quasi-steady-state ensemble defined by $\psi^{(1)}_a$, and only then slow relaxes back to the steady state $\psi^{\text{(WT)}}_a$.

\indent Consider now the effect of mutations. A single mutation in the protein could alter the energy of each conformation, leading in general to a change in the transition rate matrix $M_{ab} \to M^{\prime}_{ab} = M_{ab} + \epsilon\delta M_{ab}$, where we have used the book-keeping variable $\epsilon$ to signify a small perturbation. In matrix notation, the new steady state $\vec{\psi}^{\text{(perturb)}} = \vec{\psi}^{\text{(WT)}} + \epsilon\delta \vec{\psi}$ is, to first order in $\epsilon$:
\be
\delta\vec{\psi} = - M^{-1} \, \l \delta M \vec{\psi}^{\text{(WT)}} \r \nn \\
\ee
\noindent where $M^{-1}$ is the pesudo-inverse of $M$; more precisely, if the eigendecomposition of $M$ is $Q \Lambda Q^{-1}$, where the columns of $Q$ are the eigenvectors $\vec{\psi}^{(p)}$ and $\Lambda$ is a diagonal matrix with the eigenvalues $\lambda_p$, then:
\be
M^{-1} = \sum_{p \geq 1} \f{1}{\lambda_p} \vec{\psi}^{(p)} \, \vec{u}_p^T
\ee
\noindent where $\vec{u}_p$ is the $p$th row of the matrix $Q^{-1}$

\indent Using the expansion of $M^{-1}$ into eigenmodes, we have
\bea
\delta\vec{\psi} &=& - \l \sum_{\text{modes }p \geq 1} \f{1}{\lambda_p} \vec{\psi}^{(p)} \,  \l \vec{\psi}^{(p)} \cdot \delta M \vec{\psi}^{\text{(WT)}} \r \r \nn \\
&\approx & \f{1}{\lambda_1}\vec{\psi}^{(1)} \, \underbrace{\l - \vec{\psi}^{(1)} \cdot \delta M \vec{\psi}^{\text{(WT)}} \r}_{\phi}
\eea
\noindent where, in analogy with Eq.\ref{eq:deformingPhi}, we have defined the mutation-induced perturbation of the ensemble, $\phi$. Thus, just as mutations in a mechanical system actuate the low-energy, soft mechanical mode, mutations alter the ensemble of protein conformations along the direction of the slowest relaxational modes of the ensemble. Consequently, as mutations only sample the low-dimensional space defined by $\psi^{(1)}$, the function $F(\vec{\psi})$ reduces to a function of only the scalar variable $\phi$, $F(\vec{\psi}) = g(\phi)$. Then, writing the effect of multiple mutations $i$, $j$ on the transition rates as $\delta M = \delta_i M + \delta_j M + ...$, we once again obtain global epistasis, with $F(\vec{\psi}) = g(\phi)$, where $\phi = -\sum \vec{\psi}^{(1)}\cdot\delta_i M \vec{\psi}_0$ is the cumulative perturbation of the system to due the mutations $i$; see derivation in Sec. 1 for details and for derivation of low-rank epistasis.

\section{Global and low-rank epistasis in regulatory networks with a slow manifold}

\subsection{Theory}

\indent Generalising from models of single proteins, we now discuss epistasis in complex genetic, metabolic, or signalling regulatory networks. In each of these systems, the chemical state is specified by a set of expression levels or abundances $\mbf{n}$, with an abundance for each of the constituents of the system. In analogy to the molecular coordinates or conformational states of a protein, this a extremely high-dimensional configuration space which collectively specifies the biological function $F$ performed by the network: $F(\mbf{n})$. Examples include the net metabolic flux through a cell, or the growth rate as a function of the proteome.

\indent As before, we shall show that if mutations can be made to perturb the steady state of the network along only one dimension, then the resultant epistasis is global and, in a small-perturbation approximation, low-rank.

\indent We begin with a general dynamical system which describes the dynamics of expression levels $\mbf{n}(t)$ of $P$ components,
\be
\f{\p}{\p t}\mbf{n}(t) = \mbf{f}(\mbf{n})
\ee
\noindent where the (vector) function $\mbf{f}(\mbf{n})$ contains terms that describe postive or negative regulation, as well as production and degradation terms. We suppose only that $\mbf{f}(\mbf{n})$ depends only on the instantaneous expression level $\mbf{n}(t)$, and does not have any explicit time-dependence of its own.

\indent Denoting by $\mbf{f}_0$ the dynamics of the wildtype regulatory network, the wildtype reaches a steady-state satisfying $\mbf{f}_0(\mbf{n}_0) = 0$. The dynamics close to this steady state is described by the Jacobian $\mathcal{J}_0 \equiv \grad \mbf{f}(\mbf{n}_0)$, which plays a role analogous to the Hessian of an elastic network. In particular, the eigenvectors of the Jacobian, $\mathcal{J}_0$, $\hat{v}_m$ ($m \in \{0,...,P-1 \}$), describe the relaxational modes of the system. The eigenvalues $\lambda_m$ determine the timescales of relaxation of each mode.

\indent Analogous to a soft mechanical mode is a slow relaxational mode, defined by $\vert \lambda_0 \vert \ll \lambda_m$ for $m > 0$, associated with a \textit{slow manifold} defined by the eigenvector $\hat{v}_0$. Slow modes describe the long-time dynamics of relaxation to the steady state -- an initially perturbed system will relax quickly to the slow manifold, and only then relax slowly back to the steady state with timescale $-1/\lambda_0$. See \cite{Kaneko2015,Furusawa2018,Kaneko2018} for a description of how slow manifolds could arise in evolved regulatory networks.

\indent Note that, if there is a single slow mode, $\lambda_0$ is by necessity real. Therefore, while we shall use $\lambda_m$ in this section to denote the (potentially complex) eigenvalue itself, in the main text and elsewhere in the SI we shall use $\lambda_m$ to denote only its real part (the so-called \textit{Lyapunov exponent}).

\indent Mutations, or sustained external perturbations such as a change in nutrients, modify the regulatory dynamics $\mbf{f}_0 \l \mbf{n} \r \to \mbf{f}_0 \l \mbf{n} \r + \delta \mbf{f}(\mbf{n})$. We shall suppose that multiple mutations $i$ and $j$ alter the regulatory dynamics as $\delta \mbf{f} = \delta_i \mbf{f} + \delta_j \mbf{f}$, as would be expected for mutations that perturb different components or physical processes of a network \cite{Segre2005,Chure2019}. The new steady state $\mbf{n}_0 + \delta \mbf{n}$ satisfies
\be
\mbf{f}_0\l \mbf{n}_0 + \epsilon\,\delta \mbf{n} \r + \epsilon\, \delta \mbf{f}\l \mbf{n}_0 + \epsilon \, \delta\mbf{n} \r = 0
\ee
\noindent where we have introduced the book-keeping variable $\epsilon$ to indicate that the perturbation $\delta \mbf{f}$ is small. To first order in $\epsilon$:
\be
\delta\mbf{n} \approx - \mathcal{J}_0^{-1} \delta \mbf{f}\l \mbf{n}_0 \r \approx \f{1}{\lambda_0} \underbrace{\l - \hat{v}_0 \cdot \delta \mbf{f} \r}_{\phi} + \mathcal{O} \l \f{1}{\vert\lambda_1\vert}\r
\ee
\noindent where in the last step we have invoked the existence of a slow manifold, and defined the mutation-induced perturbation $\phi$, in analogy with Eq. \ref{eq:deformingPhi} for the mechanical networks. Thus, mutations in the background of a slow manifold $\hat{v}_0$ perturb the system primarily along the slow manifold. Consequently, exactly as we have described for proteins, epistasis between mutations will be global; see derivations above for details.

\subsection{Numerical methods}

\subsubsection*{Model}

\indent We simulate a family of regulatory networks with $P = 30$ species (`genes') whose expression levels $n_a$, $a \in \{ 1,\ldots, M \}$, evolve as:
\be \label{eq:enzymeNetwork}
\f{\p}{\p t}n_a = \sum_b k_{ab} \f{n_b}{n_b + K_{ab}} - \lambda_a \, n_a
\ee
\noindent where, as described in the main text, $K_{ab}$, $k_{ab}$ parameterise the regulation $b\to a$, and $\lambda_a$ is the decay-rate for species $a$. The choice of these parameters completely specify the particular regulatory network.

\indent For simplicity, we set the decay rates of all species $a$ to $\lambda_a = 5$, which sets the units of time in the simulation. Eq. \ref{eq:enzymeNetwork} is solved by the Euler method with time-step $\Delta t = 0.01$. Given initial conditions, the steady state $\mbf{n}_0$ is found by solving the equation until the change in each species $n_a$ over a unit interval of time is less than a specified tolerance (here, $10^{-7}$).


\subsubsection*{Varying the gap in the Lyapunov spectrum}

\indent To obtain a regulatory network with a specified gap in its Lyapunov spectrum, we begin with a randomly generated regulatory network. Explicitly, we first choose which of the 435 possible regulatory edges to include in the network, by including each with a probability $p = 0.3$ (i.e., giving a directed Erdos-Renyi graph). Then, for each included regulatory edge, the values of the constants $K_{ab}$ and $k_{ab}$ is set by sampling uniformly over the unit interval, resulting in a fully specified random regulatory network. 

\indent The steady state is found by solving the equations of motion, Eq. \ref{eq:enzymeNetwork}, with initial conditions $n_a = 0.5$; diagonalising the Jacobian $\mathcal{J}_0$ around the steady state, we obtain the (sorted) Lyapunov exponents $\lambda_i$ as the negative of the real part of the eigenvalues. Our quantity is interest is the gap between the first two exponents:
\be
\Delta\lambda \equiv \f{\lambda_1}{\lambda_0}
\ee
\indent For the initially random regulatory networks, we find that $\Delta \lambda \sim 1.5$.

\indent We then use a Monte Carlo algorithm to obtain networks with varying $\Delta \lambda$. Briefly, at each step of the algorithm, a single regulatory edge $b \to a$ is chosen, and its parameters $k_{ab}$, $K_{ab}$ are perturbed by an amount uniformly distributed in $\left[ -0.2, 0.2 \right]$. Parameters are not allowed to become negative, or to grow greater than $1$. The steady state and Lyaponov spectrum of the new network is computed, and the change in parameters accepted if:

\begin{enumerate}
	\item The gap $\Delta \lambda$ increases
	\item The inverse participation ratio, defined in terms of the elements of the lowest eigenvector $v_a$ as:
	\be
	\text{ IPR } = \sum_{i=1}^{P} \l v_a v_a^{\ast} \r^2
	\ee
	\noindent decreases or is already below $2/P$, where $P = 30$ is the total number of species in the network.
\end{enumerate}

\indent The algorithm terminates when the desired mode gap is reached. We generated a total of 320 regulatory networks, the results of which (as detailed in the next section) are shown in Figure 3 in the main text.

\subsubsection*{Mutations, fitness, and epistasis}

\indent For each network, the effect of $M = 50$ single mutations (and their double mutants) is simulated, and the epistatic matrix $\Delta \Delta F_{ij}$ computed. Each mutation $i$ is taken to affect a single regulatory edge $b \to a$ -- representing, for instance, mutations in the domain of species $b$ responsible for regulating $a$. The set of $M$ mutants is chosen by computing the `regulatory strength' ${k_{ab} n_{b}}/\l {K_{ab} + n_{b}} \r$, at the wildtype steady state, for each edge and choosing the top $M$. The effect of each mutation is to increase or decrease $k_{ab}$ and $K_{ab}$ (for the regulatory edge associated with that mutation) by $10\%$; the new steady state $\mbf{n} = \mbf{n}_0 + \delta \mbf{n}$ is found by solving Eq. \ref{eq:enzymeNetwork} initialised at the wildtype fixed point $\mbf{n}_0$.

\indent We choose as a fitness function the following:
\be
F \l \mbf{n} \r = \exp \l -\f{1}{2} \l \mbf{n} - \mbf{n}_0 \r^T \mathcal{K} \, \l \mbf{n} - \mbf{n}_0 \r \r
\ee
\noindent where $\mbf{n}$ is they steady state of the mutated network, and $\mbf{n}_0$ is the wildtype steady state. The matrix $\mathcal{K}$ is chosen randomly to realise different randomly generated fitness functions (plotted as different symbols in Fig.3 of the main text).

\indent The epistatic matrix, its rank-1 approximation, and the epistatic complexity $\kappa$ is then computed just as for the elastic network simulation.

\subsection{Principal component analysis of \textit{E. coli} proteomics data}

\indent A key consequence of the existence of a slow manifold in a regulatory network is a low-dimensional response to perturbations, be they mutations or a change in external condition. We tested this with data from the experiments of \cite{Schmidt2016}, as obtained from data the repository in \cite{Barenholz2016}. The dataset consists of protein copy number in \textit{E. coli} quantifed across 17 different growth conditions (the dataset originally consisted of 19 conditions; however, we excluded two stationary phase measurements from the analysis). After excluding proteins whose copy number were below the quantitation limit in any one condition, we end up with $1971$ protein copy numbers quantified across growth conditions. We perform principal component analysis on this dataset, the results of which are plotted in orange in Fig. 3c of the main text. As a control, we randomly scrambled the copy numbers across conditions and protein labels, resulting in the grey data points in Fig. 3c in the main text.

\section{Low-rank epistasis and evolutionary ruggedness}

\indent To gauge the evolutionary implications of low-rank epistasis, we carried out a simple numerical calculation to assess the ruggedness of a fitness landscape with a tunable rank of epistasis. For simplicity, we choose a fitness landscape with only second-order epistasis:
\be \label{eq:landscape}
F\l \mbf{s} \r = \sum_i \Delta F_i \, s_i + \sum_{i>j} \Delta\Delta F_{ij} \, s_i s_j
\ee
\noindent where, as earlier, $s_i \in \{ 0, 1\}$, for $i$ between $1$ and the sequence length $L$, specifies the genotype, and $\Delta F_i$, $\Delta\Delta F_{ij}$ are the first and second order epistatic coefficients, respectively. We parameterise $\Delta\Delta F_{ij}$ as:
\be
\Delta\Delta F_{ij} = (1-\kappa) \, \theta_i\theta_j + \kappa \, J_{ij}
\ee
\noindent where $J_{ij}$ is a symmetric, full-rank matrix. The variable $\kappa$ is the epistatic complexity, comparable to the defintion in Eq. \ref{eq:kappadef}, which tunes $\Delta\Delta F_{ij}$ from rank-1 ($\kappa = 0$) to full-rank ($\kappa = 1$).

\indent We simulate greedy adaptive walks in the landscape Eq. \ref{eq:landscape}. At each step, the mutation that leads to the largest gain in fitness is chosen. The simulation is terminated when the population reaches a local fitness maximum, i.e. when all mutations are deleterious.

\indent We use a genome length of $L = 50$, and sample $\kappa$ in steps of $0.05$ between $0$ and $1$. For each $\kappa$, $100$ random fitness landscapes are generated, by sampling $\Delta F_i$, $\theta_i$, and $J_{ij}$ from a normal distribution with mean $0$ and variance $1$. For each landscape, $100$ adaptive walks are performed starting from random initial conditions, and the number of unique fitness maxima reached is recorded. This, averaged over landscapes, is plotted (along with its standard deviation), in Fig. 4 of the main text.

\bibliographystyle{unsrt}
\bibliography{epistasis.bib}